\documentclass[
  nojss]{jss}

\usepackage{thumbpdf,lmodern}

\usepackage[utf8]{inputenc}

\author{
Miguel Sanchez-Martinez\\International
Labour Organization \And Tomasz
Woźniak\\University of Melbourne
}
\title{Forecasting with Bayesian Panel Vector Autoregressions Using the
\proglang{R} Package \pkg{bpvars} \linebreak (Version 2.0)}

\Plainauthor{Miguel Sanchez-Martinez, Tomasz Woźniak}
\Plaintitle{Forecasting with Bayesian Panel Vector Autoregressions Using
the R Package bpvars Version 2.0}
\Shorttitle{Forecasting with Bayesian Panel VARs}

\Abstract{
\noindent The \proglang{R} package \pkg{bpvars} was designed to forecast
employment, unemployment, and labour market participation rates of 189
countries. However, it is generally applicable to dynamic panel data due
to the flexibility of its modelling framework and robust coding. It
includes a family of Bayesian hierarchical panel Vector Autoregressions
(VARs) that are characterised by: (i) country-specific VAR models (ii)
with their parameters' priors centred around their global counterparts,
and (iii) featuring flexible multi-level hierarchical prior
distributions (iv) with many variants of well-established in the
literature benchmark choices, and (v) four alternative specifications
including groupping of country-specific or global parameters.
A~distinguishing feature is its implementation of missing observation
treatment based on a model-coherent Bayesian approach. These models are
accompanied by Bayesian prediction, offering a wide range of possible
specifications that aim to increase forecasting precision and comply
with various reporting standards. We also implement pseudo-out-of-sample
recursive forecasting for evaluating point and density forecast
performance. The package implements model specification, estimation, and
forecasting routines, facilitating simple workflows and reproducibility,
including estimation and forecasting results summaries and
visualisations. It achieves extraordinary computational speed thanks to
the employment of frontier econometric and numerical techniques, as well
as algorithms written in \proglang{C++}.
}

\Keywords{Dynamic Panel Data, Missing Observations, Hierarchical Prior
Distributions, Pseudo-out-of-sample Forecasting, Forecast Performance
Evaluation}
\Plainkeywords{Dynamic Panel Data, Missing Observations, Hierarchical
Prior Distributions, Pseudo-out-of-sample Forecasting, Forecast
Performance Evaluation}

\Address{
    Miguel Sanchez-Martinez\\
    International Labour Organization\\
    Macro-Economic Policies and Jobs Unit\\
ILO Research Department\\
  E-mail: \email{sanchezmartinez@ilo.org}\\
  
      Tomasz Woźniak\\
    University of Melbourne\\
    Department of Economics\\
111 Barry Street\\
3053 Carlton, VIC, Australia\\
  E-mail: \email{tomasz.wozniak@unimelb.edu.au}\\
  URL: \url{https://github.com/donotdespair}\\~\\
  }

\usepackage[utf8]{inputenc} \usepackage{tikz}

\usetikzlibrary{shapes,positioning,arrows.meta}

\usepackage{float} \usepackage[none]{hyphenat} \usepackage{amsmath,amsfonts,amssymb,amsthm,pxfonts,dsfont} \usepackage{natbib} \usepackage{graphicx}  \usepackage{color} \usepackage{geometry} \usepackage{multirow} \usepackage{booktabs} \usepackage{rotating}  \usepackage{longtable} \usepackage{subfig} \usepackage{wrapfig} \usepackage{subfig}

\emergencystretch=\maxdimen \hyphenpenalty=10000 \hbadness=10000

\begin{document}

\section{Introduction}\label{sec:nontech}

\noindent Modelling dynamic panel data requires application-specific
approaches due to the data's distinctive features, including temporal
persistence and cross-sectional dependence, and often necessitates
handling missing observations. Consequently, the literature is full of
models attempting to strike a balance between the efficient extraction
of data informational content and handling the large number of
parameters that this task would require
\citep[see][Chapter 8]{Canova2007}. Bayesian inference offers a wide
range of tools for addressing these challenges
\citep[see][]{canova2013}. Its successful applications in economics and
finance provide evidence at a high level of generality and offer
improved forecasting precision
\citep[see e.g.][]{jarocinski_responses_2010}.

This paper introduces the \pkg{bpvars} package by \cite{bpvars} for
\proglang{R} \citep[][]{Rcore} for Bayesian Forecasting with Panel
Vector Autoregressions. It was developed for the International Labour
Organization for forecasting labour market outcomes, including
unemployment, employment, and labour force participation rates, for 189
countries. The model's formulation was inspired by existing
implementations of Bayesian hierarchical modelling of global data for
other United Nations agencies, which include forecasting of carbon
dioxide emissions and temperature \citep{raftery2017}, the world's
population \citep{raftery_bayesian_2014}, or fertility rates
\citep{fosdick_regional_2014,liu_bayesian_2024}. Additionally, Bayesian
inference plays an essential role in global modelling of climate
change\footnote{See \cite{sokolov_probabilistic_2009,director_probabilistic_2021}},
demographics\footnote{See \cite{gerland2014world,raftery_joint_2014,raftery_probabilistic_2023,yu_probabilistic_2023}},
and
epidemiology\footnote{See \cite{godwin_bayesian_2017,irons_estimating_2021}}.

\begin{figure}
\centering
\includegraphics[width=0.5\linewidth,height=\textheight,keepaspectratio]{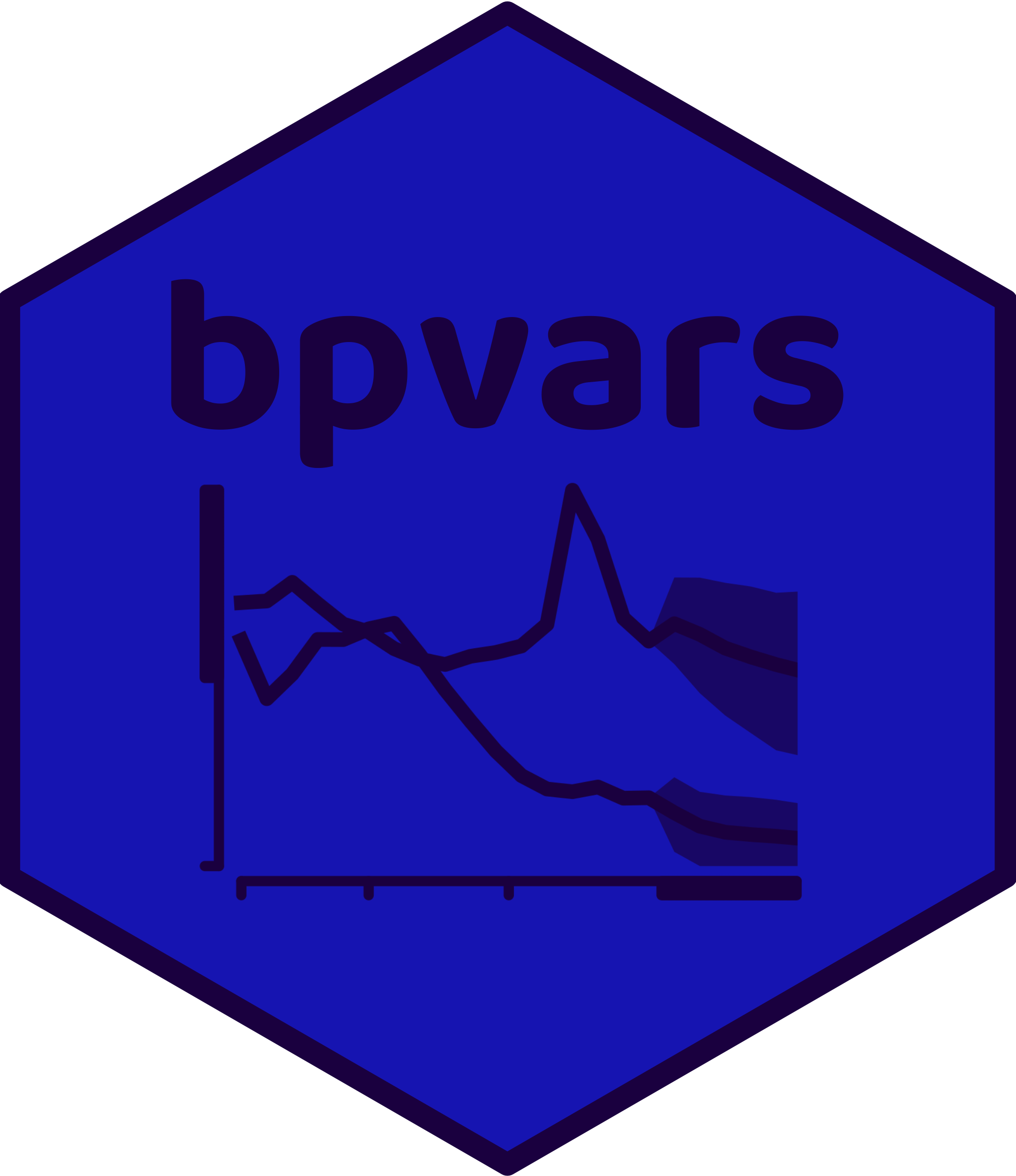}
\caption{The hexagonal package logo features unemployment rate forecasts
for Colombia and Poland that can be fully reproduced using the
\pkg{bpvars} package following a script available at:
\href{https://github.com/bsvars/hex/blob/main/bpvars/bpvars.R}{https://github.com/bsvars/hex/blob/main/bpvars/bpvars.R}}
\end{figure}

Our new family of panel Vector Autoregressions (VARs) builds on the
model proposed by \cite{jarocinski_responses_2010}. The common feature
of our models is a country-specific VAR model with autoregressive and
error term covariance parameter matrices that features prior
distributions centred around their global counterparts. In other words,
under the prior mean, the country-specific data follow a VAR model with
global parameters that are also estimated. With that respect, our
framework relates to the literature on the so-called exchangeable priors
proposed by \cite{LindleySmith1972} and random coefficient modelling
\citep[see, e.g.,][]{rendon2013}, extending them to all country-specific
model parameters in a dynamic panel modelling context. Our models treat
all variables as endogenous and contemporaneously related, capturing the
persistence of individual variables, their dynamic interactions, and
supplementing equations for individual countries with information from
other countries' data through the global prior. This is complemented by
a multiple-level hierarchical prior structure that allows the model to
adjust to various types of data without the necessity of making
arbitrary choices regarding the hyper-parameters of the prior
distributions. These features lead to a simple model formulation that
benefits from the flexibility of hierarchical priors, thereby improving
its forecasting performance.

The package includes four panel VAR models, each featuring several
variations. The benchmark model exhibits all the features mentioned
above. It can be estimated in a version proposed by
\cite{jarocinski_responses_2010}, which introduces a minimal prior
structure and the exchangeable prior for the autoregressive parameters
only. Two additional models introduce country groupings: one for
country-specific parameters and the other for global parameters. All of
these models exhibit similar hierarchical priors and feature two
alternatives for the global autoregressive parameters prior mean:
setting it to a multivariate random walk, as in the Minnesota prior by
\cite{Doan1984}, or to the pooled estimator, as proposed by
\cite{zellner_hong}. Finally, the package implements the estimation of
a~single-country VAR model, enabling users to perform comparisons with
panel counterparts with ease.

Importantly, the package provides Bayesian treatment of missing
observations. It considers three types of such observations: the initial
conditions, missing observations throughout the sampling period, and
those arising from late data releases at the end of the sample resulting
in \emph{rugged edges}. In line with the approach initiated by
\cite{guttman_bayesian_1983}, Bayesian inference treats all of these
missing observations as random variables, and estimates them alongside
with the parameters, using a likelihood-consistent joint posterior
distribution given the observed data. In other words, the missing
observations are sampled in each iteration of the estimation algorithm
from the distribution coherent with the model equations and its
distributional assumptions. Consequently, the forecasts based on a model
with missing observations accommodate the uncertainty arising from their
existence, facilitating adequate risk accountability.

To authors' best knowledge, the \pkg{bpvars} package is the first
\proglang{R} package dedicated specifically to Bayesian panel VAR
models.\footnote{The only other similar facility for handling dynamic panel data is provided by the \pkg{BGVAR} package by \cite{BGVAR,boeck_bgvar_2022} implementing Bayesian Global VAR modelling proposed by \cite{CFH2016}. Extensive code is available through \pkg{The BEAR Toolbox} by \cite{BEAR} and \cite{Dieppe2016} for \proglang{MATLAB}, offering a range of Bayesian panel VARs, including that by \cite{jarocinski_responses_2010}. We acknowledge other packages focused on frequentist estimation of models for dynamic panel data such as the \pkg{pvars} by \cite{pvars}, \pkg{panelvar} by \cite{panelvar,panelvar_pap}, and \pkg{plm} by \cite{plm,plm_pap}.}
Its distinctive feature is the combination of hierarchical modelling for
panel VARs with missing variables treatment and a wide range of
forecasting tools. This functionality is unmatched in the existing
software. Additionally, the \pkg{bpvars} package is compatible in terms
of the code structure, workflow and object design, and user experience
with the \pkg{bsvars} package by \cite{bsvars,wozniak2024} and
\pkg{bsvarSIGNs} package by \cite{WangWozniak2025,bsvarSIGNs}. This
compatibility provides an opportunity for synergies by simplifying the
learning process for the user and expanding their toolset by alternative
models. All of these packages implement frontier econometric and
numerical techniques written in \proglang{C++} facilitated using the
package \pkg{Rcpp} by \cite{Rcpp,eddelbuettel_seamless_2013} and
\cite{eddelbuettel2011rcpp}.\footnote{Similarly, linear algebra and random number generation is implemented in \proglang{C++} code using the package \pkg{RcppArmadillo} by \cite{RcppArmadillo} and \cite{eddelbuettel_rcpparmadillo2014} providing headers to the library \pkg{Armadillo} by \cite{sanderson2016armadillo}. Other similarities include using the progress bar from the package \pkg{RcppProgress} by \cite{RcppProgress}, and the structure--rich input and output out-letting using the objects from the \proglang{R} package \pkg{R6} by \cite{R6}.}

This article proceeds as follows. Section \ref{sec:hpvar} introduces our
novel benchmark model and its detailed formulation, whereas Section
\ref{sec:extensions} focuses on its extensions. Section \ref{sec:miss}
scrutinises Bayesian missing observation treatment and Section
\ref{sec:fore} presents Bayesian forecasting for dynamic panel data. The
\pkg{bpvars} package workflow detailing each of the stages of the
forecasting exercise are highlighted in Section \ref{sec:flmo}, whereas
the pseudo-out-of-sample forecasting exercise and the missing
observation treatment are outlined in Sections \ref{sec:poosf} and
\ref{sec:missing}, respectvely. Finally, Appendix \ref{sec:gibbs}
presents the detail of the estimation algorithm for the sake of
transparency and reproducibility.

\section{Hierarchical Panel Vector Autoregression}\label{sec:hpvar}

\noindent Consider a benchmark model that consists of a set of
country-specific Vector Autoregressions (VARs) with global prior
specification. In this model, labour market outcomes are forecasted with
the country-specific VAR parameters. These parameters, however,
entertain a panel model feature by sharing the same prior distribution,
which we call a global prior distribution. Such formulation of the model
can be understood as a country-specific VAR whose parameters follow a
global VAR under the prior mean. Below, the country specific model is
presented and the global prior is explained.

\subsection{Country-Specific Vector Autoregressions}\label{ssec:model}

\noindent Let an \(N\)-vector
\(\mathbf{y}_{c.t} = \begin{bmatrix}  gdp_{c.t} & ur_{c.t} & er_{c.t} & pr_{c.t} \end{bmatrix}'\)
collect the dependent variables for country \(c\) at time \(t\), where
the country indicator takes values \(c \in\{ 1,\dots,C\}\) with a total
number of \(C\) countries in the sample data, and the time indicator
\(t \in\{ 1,\dots,T_c\}\), with the country-specific sample size
\(T_c\). These variables follow a multivariate dynamic specification,
namely, the Gaussian VAR model \citep[see][]{Sims1980} given by
\begin{align}
\mathbf{y}_{c.t} &= \mathbf{A}_{c.1}\mathbf{y}_{c.t-1} + \dots + \mathbf{A}_{c.p} \mathbf{y}_{c.t-p} + \mathbf{A}_{c.d}\mathbf{d}_{c.t} + \boldsymbol\epsilon_{c.t},\label{eq:csvar}\\
\boldsymbol\epsilon_{c.t}|\mathbf{y}_{c.t-1}, \dots, \mathbf{y}_{c.t-p} &\sim iid\mathcal{N}_N\left(\mathbf{0}_N, \boldsymbol\Sigma_c\right),\label{eq:csshock}
\end{align} where \(\mathbf{A}_{c.l}\) are \(N\times N\) are
autoregressive matrices at lag \(l\), \(\mathbf{d}_{c.t}\) is the
\(D\)-vector of deterministic terms, and \(\mathbf{A}_{c.d}\) is the
\(N\times D\) matrix of the corresponding coefficients,
\(\boldsymbol\epsilon_{c.t}\) is the \(N\)-vector of error terms that is
normally distributed with covariance matrix \(\boldsymbol\Sigma_c\).
Additionally, this dynamic model relies on initial conditions
\(\mathbf{y}_{c.0}, \dots, \mathbf{y}_{c.-(p-1)}\) which are considered
in Section~\ref{sec:miss}.

In the model from expressions \eqref{eq:csvar} and \eqref{eq:csshock},
all the variables are treated as endogenous through the specification of
the joint conditional normal distribution with covariance
\(\boldsymbol\Sigma_c\) and their dynamics and temporal
inter-dependencies are captured by the autoregressive parameters
\(\mathbf{A}_{c.l}\). These features decide on the improved forecasting
performance of VAR models compared to their univariate or static
alternatives.

Rewrite this model in a matrix notation. Define a \(T_c\times N\) matrix
\(\mathbf{Y}_c = \begin{bmatrix} \mathbf{y}_{c.1} &\dots & \mathbf{y}_{c.T_c} \end{bmatrix}'\),
a \(T_c\times K\) matrix
\(\mathbf{X}_c = \begin{bmatrix} \mathbf{x}_{c.1} &\dots & \mathbf{x}_{c.T_c} \end{bmatrix}'\),
where
\(\mathbf{x}_{c.t} = \begin{bmatrix} \mathbf{y}_{c.t-1}' &\dots & \mathbf{y}_{c.t-p}' & \mathbf{d}_{c.t}' \end{bmatrix}'\),
and \(K=Np+D\), a \(T_c\times N\) matrix
\(\mathbf{E}_c = \begin{bmatrix} \boldsymbol\epsilon_{c.1} &\dots & \boldsymbol\epsilon_{c.T_c} \end{bmatrix}'\),
and a \(K\times N\) matrix
\(\mathbf{A}_c = \begin{bmatrix} \mathbf{A}_{c.1} &\dots & \mathbf{A}_{c.p} & \mathbf{A}_{c.d} \end{bmatrix}'\).
Then, the model from \eqref{eq:csvar}--\eqref{eq:csshock} can be written
in an equivalent form as \begin{align}
\mathbf{Y}_c &= \mathbf{X}_c\mathbf{A}_c + \mathbf{E}_c,\label{eq:csvarmat}\\
\mathbf{E}_c | \mathbf{X}_c &\sim\mathcal{MN}_{T_c\times N}\left(\mathbf{0}_{T_c\times N}, \mathbf{I}_{T_c}, \boldsymbol\Sigma_c\right),\label{eq:csshockmat}
\end{align} where \(\mathcal{MN}_{T_c\times N}()\) denotes a
matrix-variate normal distribution for a \(T_c\times N\) matrix
\citep[see][]{blr1999,wozniak2016}.\footnote{Let operator $\text{vec}()$ stack the columns of $\mathbf{E}_c$ one under another in a $T_cN\times 1$ vector $\text{vec}(\mathbf{E}_c)$. Then the distribution specification in \eqref{eq:csshockmat} with the mean matrix $\mathbf{0}_{T_c\times N}$, the row-specific covariance parameter $\boldsymbol\Sigma_c$, and the column-specific covariance parameter $\mathbf{I}_{T_c}$, is equivalent to the multivariate normal distribution $\text{vec}(\mathbf{E}_c)\sim\mathcal{N}_{T_cN}\left(\mathbf{0}_{T_cN\times1}, \boldsymbol\Sigma_c\otimes\mathbf{I}_{T_c}\right)$, where $\otimes$ denotes the Kronecker product of two matrices.}
Note that the assumptions in matrix specification~\eqref{eq:csshockmat}
correspond to those from assumption \eqref{eq:csshock}. It is further
used to introduce the prior structure for the model.

\subsection{Global Hierarchical Prior Distribution}\label{ssec:gprior}

\noindent Bayesian inference and estimation requires the specification
of prior distributions of the model's parameters. These distributions
are specified based on the feasibility of estimation, interpretability,
and to optimise forecasting performance. In what follows, the complete
prior specification is presented for this benchmark model and the
corresponding estimation procedure is explained in
Appendix~\ref{sec:gibbs}.

The main objectives motivating the prior distributions selection are:

\begin{itemize}
\item to ensure flexible prior specification for the country-specific parameters allowing them to vary substantially for different countries,
\item to facilitate the estimation of the global parameters, thereby giving the model a panel data model interpretation,
\item to grant the prior distribution the interpretability of the Minnesota prior or pooled estimator, which has been proven to improve forecasting performance for macroeconomic aggregates and panel data models,
\item to allow the data to flexibly determine the level of prior shrinkage,
\item to result in efficient Bayesian estimation through Gibbs sampler.
\end{itemize}

The prior distribution for the country-specific parameters
\(\mathbf{A}_c\) and \(\boldsymbol\Sigma_c\) is specified in a
hierarchical manner, which allows for the estimation of the prior
hyper-parameters addressing the objectives stated above. Its main
characteristic is the interpretation of the prior means for
country-specific autoregressive parameters
\(\mathbb{E}\left[\mathbf{A}_c\right]=\mathbf{A}\) and error term
covariance matrix
\(\mathbb{E}\left[\boldsymbol\Sigma_c\right]=\boldsymbol\Sigma\) as
global parameters, that is, invariant over countries \(c\). These prior
expected values imply a VAR model with global parameters is given by:
\begin{align}
\mathbf{Y}_c &= \mathbf{X}_c\mathbf{A} + \mathbf{E}_c,\label{eq:gvar}\\
\mathbf{E}_c | \mathbf{X}_c &\sim\mathcal{MN}_{T_c\times N}\left(\mathbf{0}_{T_c\times N},\mathbf{I}_{T_c}, \boldsymbol\Sigma\right).\label{eq:gshock}
\end{align} This prior specification is implemented by assuming a
convenient matrix-variate normal inverse Wishart distribution
\citep[see][]{karlsson2013,wozniak2016} given by \begin{align}
\mathbf{A}_c, \boldsymbol\Sigma_c | \mathbf{A}, \mathbf{V}, \mathbf{\Sigma}, \nu &\sim\mathcal{MNIW}_{K\times N}\left(\mathbf{A}, \mathbf{V}, (N - \nu - 1)\mathbf{\Sigma}, \nu\right)\label{eq:csmnivprior}
\end{align} that additionally includes hyper-parameters determining the
scale and shape of the prior distribution for \(\mathbf{A}_c\) and
\(\boldsymbol\Sigma_c\), namely, a \(K\times K\) column-specific
covariance matrix \(\mathbf{V}\) and the shape parameter \(\nu\).
Additional advantage of this specification is that it leads to a
convenient Gibbs sampler for the estimation of the model.

The global parameters are estimated, which is facilitated by Bayesian
hierarchical modelling. Therefore, the global autoregressive matrix,
\(\mathbf{A}\), follows a matrix-variate normal distribution with the
mean \(K\times N\) matrix \(m\underline{\mathbf{M}}\), \(K\times K\)
column-specific covariance \(\mathbf{V}\), and \(N\times N\)
row-specific covariance \(s\underline{\mathbf{S}}\) denoted by

\begin{align}
\mathbf{A} \mid \mathbf{V}, m, s &\sim\mathcal{MN}_{K\times N}\left(m\underline{\mathbf{M}}, \mathbf{V}, s\underline{\mathbf{S}}\right),\label{eq:priorA}
\end{align} where \(\underline{\mathbf{M}}\) and
\(\underline{\mathbf{S}}\) are fixed matrices of appropriate sizes and
types while scalar hyper-parameters \(m\) and \(s\) are further
estimated.

The global error term covariance matrix, \(\mathbf{\Sigma}\), follows a
Wishart distribution with \(N\times N\) scale matrix
\(s\underline{\mathbf{S}}_\Sigma\) and shape parameter
\(\underline{\mu}_\Sigma\) \begin{align}
\mathbf{\Sigma}\mid s, \nu &\sim\mathcal{W}_{N}\left(s\underline{\mathbf{S}}_\Sigma,\underline{\mu}_\Sigma\right),\label{eq:priorSigma}
\end{align} where the matrix \(\underline{\mathbf{S}}_\Sigma\) and shape
parameter \(\underline{\mu}_\Sigma\) are fixed, while the positive
scalar \(s\) is estimated.

\subsection{Hierarchical Prior Distribution}\label{ssec:hprior}

\noindent In the hierarchical Panel VAR model proposed here, all of the
hyper-parameters of the prior in \eqref{eq:csmnivprior} are estimated.
This gives the model the advantage of fitting the data closely, while
avoiding the necessity of making arbitrary choices regarding the values
of these hyper-parameters.

Consequently, a prior distribution is assumed for the column-specific
covariance \(\mathbf{V}\) that controls the level of shrinkage of the
autoregressive parameters \(\mathbf{A}_c\) and \(\mathbf{A}\) around
their respective prior means. It is set to the inverse-Wishart
distribution with scale \(w\underline{\mathbf{W}}\) and shape
\(\underline{\eta}\) \begin{align}
\mathbf{V} \mid w &\sim\mathcal{IW}_{N}\left(w\underline{\mathbf{W}}, \underline{\eta}\right),\label{eq:priorV}
\end{align} with the \(K\times K\) scale matrix
\(\underline{\mathbf{W}}\) and the shape parameter \(\underline{\eta}\)
being fixed and the positive scalar \(w\) estimated. The shape parameter
\(\nu\) follows an exponential distribution with mean
\(\underline\lambda\) denoted by \begin{align}
\nu &\sim\exp\left(\underline\lambda\right).\label{eq:priorNU}
\end{align} The prior specification is complemented by the average
global persistence hyper-parameters \(m\), and scaling factors \(w\) and
\(s\) following the normal, gamma, and inverted gamma 2 prior
distributions respectively \begin{align}
m &\sim\mathcal{N}\left(\underline{\mu}_m, \underline{\sigma}_m^2\right),\\
w &\sim\mathcal{G}\left(\underline{s}_w, \underline{a}_w\right),\\
s &\sim\mathcal{IG}2\left(\underline{s}_s, \underline{\nu}_s\right).\label{eq:sprior}
\end{align}

To summarise, the joint prior distribution for the parameters of the
model is given by \begin{multline}
p\left(\mathbf{A}_c, \boldsymbol\Sigma_c, \mathbf{A}, \mathbf{V}, \mathbf{\Sigma}, \nu, m, w, s\right) =
p\left(\mathbf{A}_c, \boldsymbol\Sigma_c | \mathbf{A}, \mathbf{V}, \mathbf{\Sigma}, \nu\right)
p\left(\mathbf{A} \mid \mathbf{V}, m, s\right)
p\left(\mathbf{\Sigma}\mid s\right)\\
\times 
p\left(\mathbf{V} \mid w\right)
p(\nu)p(m)p(w)p(s),
\end{multline} where the particular distributions are as follows:
\begin{align}
\mathbf{A}_c, \boldsymbol\Sigma_c | \mathbf{A}, \mathbf{V}, \mathbf{\Sigma}, \nu &\sim\mathcal{MNIW}_{K\times N}\left(\mathbf{A}, \mathbf{V}, (N - \nu - 1)\mathbf{\Sigma}, \nu\right)\\
\mathbf{A} \mid \mathbf{V}, m, s &\sim\mathcal{MN}_{K\times N}\left(m\underline{\mathbf{M}}, \mathbf{V}, s\underline{\mathbf{S}}\right)\\
\mathbf{\Sigma}\mid s &\sim\mathcal{W}_{N}\left(s\underline{\mathbf{S}}_\Sigma,\underline{\mu}_\Sigma\right)\\
\mathbf{V} \mid w &\sim\mathcal{IW}_{N}\left(w\underline{\mathbf{W}}, \underline{\eta}\right)\\
\nu &\sim\exp\left(\underline\lambda\right)\label{eq:prior_nu}\\
m &\sim\mathcal{N}\left(\underline{\mu}_m, \underline{\sigma}_m^2\right)\label{eq:prior_m}\\
w &\sim\mathcal{G}\left(\underline{s}_w, \underline{a}_w\right)\\
s &\sim\mathcal{IG}2\left(\underline{s}_s, \underline{\nu}_s\right).
\end{align}

\subsection{Fixed Prior Hyper-Parameters}\label{ssec:fphp}

These prior distributions depend on fixed hyper-parameters which in our
notation are underscored. In what follows, we provide a justification
for their default values, which are then utilized in the \pkg{bpvars}
package.

The package offers two alternative values of the matrix contributing to
the global autoregressive parameters prior mean,
\(\underline{\mathbf{M}}\). The first choice is motivated by the
interpretability of the Minnesota Prior proposed by \cite{Doan1984}, in
which case this matrix is set to
\(\begin{bmatrix} \mathbf{I}_N & \mathbf{0}_{N\times K-N} \end{bmatrix}'\).
It implies that the prior mean of the own lag for each of the variables
is estimated by the value of another hyper-parameter \(m\)
pre-multiplying this matrix in \eqref{eq:priorA}. The alternative choice
draws on the idea by \cite{zellner_hong} where this matrix is set to the
polled estimator equal to
\(\left( \sum_{c=1}^{C}\mathbf{X}_c'\mathbf{X}_c \right)^{-1}\left( \sum_{c=1}^{C}\mathbf{X}_c'\mathbf{Y}_c \right)\).
The prior mean specification is complemented by a normal prior
distribution for the hyper-parameter \(m\) with mean
\(\underline{\mu}_m = 1\) and the variance
\(\underline{\sigma}_m^2 = 1\). This distribution centres the prior
around the matrix \(\underline{\mathbf{M}}\), that is around random walk
process in the Minnesota prior, and the pooled estimator otherwise.

The \(\mathbf{V}\) parameter prior scale matrix is set to
\(\underline{\mathbf{W}} = \text{diag} \begin{pmatrix} \boldsymbol\imath_N \otimes \mathbf{p}^{-2}& 100 \end{pmatrix}\),
where \(\mathbf{p}\) is a \(p\)-vector of values from 1 to \(p\),
implements the feature of the Minnesota prior for global parameters
where the shrinkage towards the prior mean becomes exponentially
stronger for autoregressive matrices \(\mathbf{A}_i\) with increasing
lag order \(i = 1,\dots,p\). The estimated hyper-parameter
pre-multiplying this matrix, namely \(w\), features a gamma prior with
the scale \(\underline{s}_w = 1\) and shape \(\underline{a}_w = 1\).
These values make the prior distribution little informative and lets the
data decide on the underlying estimate. The row-specific covariance
matrix of \(\mathbf{A}\) is a product of the identity matrix
\(\underline{\mathbf{S}} = \mathbf{I}_N\) and the estimated
hyper-parameter \(s\), featuring the inverted gamma 2 prior with the
scale and shape set to \(\underline{s}_s = 1\) and
\(\underline{\nu}_s = 3\) respectively.

Similar choices are made for the prior scale of the global covariance
matrix \(\mathbf{\Sigma}\) being the identity matrix
\(\underline{\mathbf{S}}_\Sigma = \mathbf{I}_N\) pre-multiplied by the
estimated \(s\). The shape parameter of this Wishart distribution is set
to \(\underline{\mu}_\Sigma = N + 1\), which ensures finite prior
variance of \(\mathbf{\Sigma}\). Similarly, the value of the shape
parameter for the prior distribution in \eqref{eq:priorV} is set to
\(\underline{\eta} = N + 1\).

Finally, the exponential prior for the degrees of freedom parameter
\(\nu\) is set to \(\underline\lambda = 72\), which assigns 50\% of the
prior probability to the degrees of freedom parameter being less than
50. This choice makes the prior span the part of the parameter space
implying Student-t like distribution for low values of \(\nu\), as well
as close approximations of the normal distribution for values of
\(\nu>30\).

\section{Model Extensions}\label{sec:extensions}

\subsection{A Model by Jaroci\'nski}\label{ssec:jaro}

The benchmark specification presented in Section \ref{sec:hpvar} has an
alternative prior specification proposed by
\cite{jarocinski_responses_2010}. This model facilitates the estimation
of the country-specific parameters and the global autoregressive matrix
assuming a minimal prior structure. It features equations
\eqref{eq:csvarmat} and \eqref{eq:csshockmat} with the prior
distribution for the country-specific parameters given by \begin{align}
\mathbf{A}_c | \boldsymbol\Sigma_c, \mathbf{A}, s &\sim\mathcal{MN}_{K\times N}\left(\mathbf{A}, s\underline{\mathbf{W}}, \mathbf{\Sigma}_c\right)\label{eq:jaropriorA}\\
\boldsymbol\Sigma_c &\propto \det\left(\boldsymbol\Sigma_c\right)^{-\frac{N + 1}{2}}\label{eq:jaropriorS}
\end{align} In this model, the global autoregressive parameters follow
an improper prior distribution: \begin{align}
p\left(\mathbf{A}\right) &\propto 1,\label{eq:jaropriorAglobal}
\end{align} and \(s\) follows the inverted gamma 2 prior distribution as
in \eqref{eq:sprior} with \(\underline{s}_s=\underline{\nu}_s= 0.001\).

\subsection{A Model with Country Grouping}\label{ssec:specCG}

As a variation on the country-specific VAR model, we consider a model
with country grouping. Consider a set of \(C\) countries grouped into
\(G\) groups, where each group \(g\in\{1,\dots,G\}\) contains \(C_g\)
countries. Each group \(g\) has its own parameters, \(\mathbf{A}_{g}\)
and \(\boldsymbol\Sigma_{g}\), defining the VAR model with country
grouping for country \(c\) from group \(g\) given by \begin{align}
\mathbf{Y}_c &= \mathbf{X}_c\mathbf{A}_g + \mathbf{E}_c,\label{eq:csvarmatg}\\
\mathbf{E}_c | \mathbf{X}_c &\sim\mathcal{MN}_{T_c\times N}\left(\mathbf{0}_{T_c\times N}, \mathbf{I}_{T_c}, \boldsymbol\Sigma_g\right).\label{eq:csshockmatg}
\end{align} In other words, this model is specified by imposing
restrictions on the country-specific parameters such that
\(\mathbf{A}_c = \mathbf{A}_g\) and
\(\boldsymbol\Sigma_c = \boldsymbol\Sigma_g\). The group allocations can
be fixed and determined, for example, by geographical location or
economic development, or they can be estimated from the data given a
pre-specified number of groups \(G\).

The group-specific parameters follow the hierarchical prior distribution
given by \begin{align}
\mathbf{A}_g, \boldsymbol\Sigma_g | \mathbf{A}, \mathbf{V}, \mathbf{\Sigma}, \nu &\sim\mathcal{MNIW}_{K\times N}\left(\mathbf{A}, \mathbf{V}, (N - \nu - 1)\mathbf{\Sigma}, \nu\right),\label{eq:gsmnivpriorg}
\end{align} that is complemented by the hierarchical structure as
presented in in Section \ref{sec:hpvar}.

\subsection{A Model with Global Prior Grouping}\label{ssec:specGPG}

Another variation is a model featuring equations \eqref{eq:csvarmat} and
\eqref{eq:csshockmat} in which the country-specific parameters have
group-specific global parameters. This model specification is
implemented by setting the prior expectations to
\(\mathbb{E}\left[\mathbf{A}_c\right]=\mathbf{A}_g\) and
\(\mathbb{E}\left[\boldsymbol\Sigma_c\right]=\boldsymbol\Sigma_g\) and
to let the country groupings to be fixed and specified by the user or
estimated for a fixed group number \(G\). Consequently, the
country-specific parameters follow the hierarchical prior distribution
given by \begin{align}
\mathbf{A}_c, \boldsymbol\Sigma_c | \mathbf{A}_g, \mathbf{V}, \mathbf{\Sigma}_g, \nu &\sim\mathcal{MNIW}_{K\times N}\left(\mathbf{A}_g, \mathbf{V}, (N - \nu - 1)\mathbf{\Sigma}_g, \nu\right)\label{eq:csmnivpriorgpg}
\end{align} with the priors for the group-specific global parameters
given by: \begin{align}
\mathbf{A}_g \mid \mathbf{V}, m, s &\sim\mathcal{MN}_{K\times N}\left(m\underline{\mathbf{M}}, \mathbf{V}, s\underline{\mathbf{S}}\right),\label{eq:priorAgpg}\\
\mathbf{\Sigma}_g\mid s, \nu &\sim\mathcal{W}_{N}\left(s\underline{\mathbf{S}}_\Sigma,\underline{\mu}_\Sigma\right).
\end{align} Similarly, to other specifications, the remaining prior
hierarchy is as described in Section \ref{sec:hpvar} with the same
choices for the values of matrix \(\underline{\mathbf{M}}\).

\subsection{Vector Autoregressions for Individual Countries}\label{ssec:specvars}

Finally, the package allows the estimation of VAR models for individual
countries. This facility is provided for comparisons with the
Hierarchical Panel models. In this model, the country specific
parameters are estimated independently for each country \(c\) by setting
the model equations as in \eqref{eq:csvarmat} and \eqref{eq:csshockmat},
and assuming the following prior distribution: \begin{align}
\mathbf{A}_c, \boldsymbol\Sigma_c | m, s, w, \nu &\sim\mathcal{MNIW}_{K\times N}\left(m\underline{\mathbf{M}}, w\underline{\mathbf{W}}, s\underline{\mathbf{S}}, \nu\right),\label{eq:csmnivpriorvas}
\end{align} where the hyper-parameters \(m\), \(s\), \(w\), and \(\nu\)
may be pre-specified, or estimated. In the latter case, the
hyper-parameters \(m\) and \(\nu\) follow the distributions specified in
\eqref{eq:prior_m} and \eqref{eq:prior_nu}, respectively, while those
for \(s\) and \(w\) are: \begin{align}
w &\sim\mathcal{IG}2\left(\underline{s}_w, \underline{\nu}_w\right)\\
s &\sim\mathcal{G}\left(\underline{s}_s, \underline{a}_s\right).
\end{align}

An alternative prior specification for these country-specific models is
the diffuse prior set as: \begin{align}
p\left(\mathbf{A}_c, \boldsymbol\Sigma_c\right) \propto \det\left(\boldsymbol\Sigma_c\right)^{-\frac{N + 1}{2}},
\end{align} ensuring that the posterior mean estimator for the
country-specific parameters is equal to the corresponding maximum
likelihood estimator \citep[see][]{karlsson2013}.

\section{Bayesian Missing Observation Treatment}\label{sec:miss}

Dynamic panel data often contain missing observations due to varying
starting periods of data collection in various jurisdictions, temporary
breaks in data collection, sluggish data announcements, and other
reasons. The \pkg{bpvars} package comprehensively implements solutions
to two aspects of the problem. Firstly, it takes advantage of all of the
provided observations by specifying the likelihood function for all of
them and estimating the initial values
\(\mathbf{y}_{c.0}, \dots, \mathbf{y}_{c.-(p-1)}\). Secondly, it treats
the missing observations within the sample period as additional
parameters to be estimated. Consequently, Bayesian inference specifies
the joint distribution for the initial conditions, missing and observed
data based on the parametric assumptions of the model and estimates the
initial conditions and missing data along with the model parameters by
using their implied conditional distribution given the observed data.
This sampler is integrated into the Gibbs sampler presented in Appendix
\ref{sec:gibbs}. All the models in the \pkg{bpvars} package estimate the
initial values, whereas the missing observations treatment is
automatically applied if the provided data include missing values.

Begin by rewriting the model in a convenient form. Consider a vector
collecting the initial conditions and data, observed and missing,
\(\mathbf{y}_c\) of dimension \(T_c + p\). It stacks the
country-specific vectorised data
\(\mathbf{y}_c = \begin{bmatrix} \mathbf{y}_{c.-(p-1)}' &\dots & \mathbf{y}_{c.0}' & \mathbf{y}_{c.1}' &\dots & \mathbf{y}_{c.T_c}' \end{bmatrix}'\).
Let the vector of initial conditions and missing observations,
\(\mathbf{y}_{c.m}\), be created using a \(T_{c.m}\times T_c+p\)
selection matrix \(\mathbf{S}_{c.m}\), such that
\(\mathbf{y}_{c.m} = \mathbf{S}_{c.m}\mathbf{y}_c\). Similarly, let the
vector of observed data, \(\mathbf{y}_{c.o}\), be created using a
\(T_{c.o}\times T_c+p\) selection matrix \(\mathbf{S}_{c.o}\), such that
\(\mathbf{y}_{c.o} = \mathbf{S}_{c.o}\mathbf{y}_c\).

The joint distribution of the initial conditions, missing and observed
data, \(\mathbf{y}_c\), is based on rewritten model equations
\eqref{eq:csvar} and \eqref{eq:csshock} that take the form:
\begin{align}
\mathbf{H}_{\mathbf{A}_c}\mathbf{y}_c &= \boldsymbol\mu_c + \boldsymbol\epsilon_c,\label{eq:missmodel}\\
\boldsymbol\epsilon_c &\sim \mathcal{N}_{T_c+p}\left(\mathbf{0}_T, \text{bdiag}\left(\mathbf{I}_p \otimes \boldsymbol\Sigma, \mathbf{I}_{T_c} \otimes \boldsymbol\Sigma_c\right)\right),\label{eq:missshock}
\end{align} where \(\text{bdiag}\) constructs a block-diagonal matrix
from the provided blocks. For simplicity of exposition we present the
notation for \(p=1\). Vector
\(\boldsymbol\mu_c = \begin{bmatrix} (\mathbf{A}_{d}\mathbf{d}_{c.0})'&  (\mathbf{A}_{c.d}\mathbf{d}_{c.1})' &\dots & (\mathbf{A}_{c.d}\mathbf{d}_{c.T_c})' \end{bmatrix}'\)
stacks the deterministic components, and
\(\epsilon_c = \begin{bmatrix} \boldsymbol\epsilon_{c.0}' & \boldsymbol\epsilon_{c.1}' &\dots & \boldsymbol\epsilon_{c.T_c}' \end{bmatrix}'\)
stacks the error terms. The square matrix of order \(N(T_c+p)\),
\(\mathbf{H}_{\mathbf{A}_c}\), is constructed based on the
autoregressive parameters such that: \begin{align}
\mathbf{H}_{\mathbf{A}_c} = \begin{bmatrix}
-\mathbf{A}_1 & \mathbf{I}_N & \mathbf{0}_{N\times N} & \dots& \dots & \mathbf{0}_{N\times N}\\
-\mathbf{A}_{c.1} & \mathbf{I}_N & \mathbf{0}_{N\times N} & \dots& \dots & \mathbf{0}_{N\times N}\\
\mathbf{0}_{N\times N} &-\mathbf{A}_{c.1} & \mathbf{I}_N  & \mathbf{0}_{N\times N}& \dots & \mathbf{0}_{N\times N}\\
\vdots & \vdots & \vdots & \ddots & \ddots & \vdots\\
\mathbf{0}_{N\times N} & \dots & \dots & \mathbf{0}_{N\times N} &-\mathbf{A}_{c.1} & \mathbf{I}_N 
\end{bmatrix}.
\end{align} The first row of matrix \(\mathbf{H}_{\mathbf{A}_c}\) is
specified for the initial condition \(\mathbf{y}_0\) in line with the
global prior assumption in equation~\eqref{eq:gvar}. Similarly, the
first element of \(\boldsymbol\mu_c\) and the top-left-hand-side block
of the covariance matrix in~\eqref{eq:missshock} include global
parameters. The remaining rows of matrix \(\mathbf{H}_{\mathbf{A}_c}\)
rely on country-specific parameters in line with the model equations.

The resulting joint distribution of the initial conditions, missing and
observed data, \(\mathbf{y}_c\), is the following multivariate normal
distribution: \begin{align}
\mathbf{y}_c \mid \mathbf{A}, \mathbf{A}_c, \boldsymbol\Sigma, \boldsymbol\Sigma_c &\sim \mathcal{N}_{T_c+p}\left(\overline{\mathbf{y}}_c, \boldsymbol\Sigma_{\mathbf{y}_c}\right),\label{eq:jointmiss}\\
\overline{\mathbf{y}}_c &= \mathbf{H}_{\mathbf{A}_c}^{-1}\boldsymbol\mu_c,\\
\boldsymbol\Sigma_{\mathbf{y}_c} &= \mathbf{H}_{\mathbf{A}_c}^{-1}\text{bdiag}\left(\mathbf{I}_p \otimes \boldsymbol\Sigma, \mathbf{I}_{T_c} \otimes \boldsymbol\Sigma_c\right)\mathbf{H}_{\mathbf{A}_c}^{-1\prime}.
\end{align} It is subsequently used to provide a conditional
distribution of the missing values given the observed data:
\begin{align}
\mathbf{y}_m \mid \mathbf{y}_o, \mathbf{A}, \mathbf{A}_c, \boldsymbol\Sigma, \boldsymbol\Sigma_c &\sim \mathcal{N}_{T_{c.m}}\left(\overline{\mathbf{y}}_m, \boldsymbol\Sigma_{\mathbf{y}_m}\right),\label{eq:condmiss}\\
\overline{\mathbf{y}}_m &= \mathbf{S}_m\overline{\mathbf{y}}_c + \mathbf{S}_m\boldsymbol\Sigma_{\mathbf{y}_c}\mathbf{S}_o'\left(\mathbf{S}_o\boldsymbol\Sigma_{\mathbf{y}_c}\mathbf{S}_o'\right)^{-1}\left(\mathbf{y}_o - \mathbf{S}_o\overline{\mathbf{y}}_c\right),\\
\boldsymbol\Sigma_{\mathbf{y}_m} &= \mathbf{S}_m\boldsymbol\Sigma_{\mathbf{y}_c}\mathbf{S}_m' - \mathbf{S}_m\boldsymbol\Sigma_{\mathbf{y}_c}\mathbf{S}_o'\left(\mathbf{S}_o\boldsymbol\Sigma_{\mathbf{y}_c}\mathbf{S}_o'\right)^{-1}\mathbf{S}_o\boldsymbol\Sigma_{\mathbf{y}_c}\mathbf{S}_m'.
\end{align} A sampler from the distribution~\eqref{eq:condmiss} is used
in every iteration of the Gibbs sampler, and the model's parameters are
sampled conditionally on the observed data and the current draw of the
initial values and missing values.

\section{Bayesian Forecasting}\label{sec:fore}

\noindent Bayesian forecasting in this panel VAR model is performed by
sampling from its predictive density. In this section, the conditional
predictive density -- that is, the common element with frequentist
approach -- is derived, a numerical sampling procedure from this
distribution is presented, and a range of techniques, including
marginal, restricted, and conditional forecasting are discussed. Then
the pseudo-out-of-sample forecasting is explained and complemented by
the measures of forecast accuracy.

\subsection{Conditional Predictive Density}\label{ssec:cpd}

\noindent The joint conditional predictive density of the unknown future
values to be predicted at the forecast horizon \(h\) from 1 to \(H\)
periods ahead, given the forecast origin at period \(T_c\), denoted by
\(\mathbf{y}_{c.T_c + H}, \dots, \mathbf{y}_{c.T_c + 1}\), is formed on
the basis of the Bayesian Hierarchical Panel Vector Autoregressive
model. This joint conditional density is factorised in terms of the
one-period-ahead conditional densities \begin{align*}
p\left(\mathbf{y}_{c.T_c + H}, \dots, \mathbf{y}_{c.T_c + 1} \mid \mathbf{Y}_c, \mathbf{X}_c, \mathbf{A}_c, \boldsymbol\Sigma_c\right) =& 
p\left(\mathbf{y}_{c.T_c + H} \mid \mathbf{y}_{c.T_c + H - 1}, \dots, \mathbf{y}_{c.T_c + 1}, \mathbf{Y}_c, \mathbf{X}_c, \mathbf{A}_c, \boldsymbol\Sigma_c\right)\\
&\times\dots\\
&\times p\left(\mathbf{y}_{c.T_c + 2} \mid \mathbf{y}_{c.T_c + 1}, \mathbf{Y}_c, \mathbf{X}_c, \mathbf{A}_c, \boldsymbol\Sigma_c\right)\\
&\times p\left(\mathbf{y}_{c.T_c + 1} \mid \mathbf{Y}_c, \mathbf{X}_c, \mathbf{A}_c, \boldsymbol\Sigma_c\right).
\end{align*} Notably, each of the conditional predictive densities on
the RHS of the expression above is a one-period-ahead forecasting
density determined by the model in \eqref{eq:csvar} and
\eqref{eq:csshock} that can be presented as \begin{align}
p\left(\mathbf{y}_{c.t+1} \mid \mathbf{y}_{c.t}, \dots, \mathbf{y}_{c.t-p+1}, \mathbf{A}_c, \boldsymbol\Sigma_c\right) = \mathcal{N}_N\left(
\mathbf{A}_{c.1}\mathbf{y}_{c.t} + \dots + \mathbf{A}_{c.p} \mathbf{y}_{c.t-p+1} + \mathbf{A}_{c.d}\mathbf{d}_{c.t+1} , \boldsymbol\Sigma_c
\right)\label{eq:opapredictive}
\end{align} In what follows, the density in equation
\eqref{eq:opapredictive} is used to form the Bayesian predictive density
in Section \ref{ssec:bpd} and to facilitate a straightforward numerical
algorithm to obtain the random draws from it in Section
\ref{ssec:sampling}.

\subsection{Bayesian Predictive Density}\label{ssec:bpd}

\noindent The purpose of Bayesian forecasting is to obtain a sample of
random numbers drawn from the predictive density. This density is
defined as the joint predictive density of the future unknown values,
\(\mathbf{y}_{c.T_c + H}, \dots, \mathbf{y}_{c.T_c + 1}\), given sample
data in matrices \(\mathbf{Y}_c\) and \(\mathbf{X}_c\). An important
feature is that this density does not involve conditioning on the
parameter values, which distinguishes Bayesian from frequentist
forecasting. This lack of dependence is obtained by integrating out the
parameter values from the predictive density using the posterior
distribution \begin{multline}
p\left( \mathbf{y}_{c.T_c + H}, \dots, \mathbf{y}_{c.T_c + 1} \mid \mathbf{Y}_c, \mathbf{X}_c\right) = \int p\left(\mathbf{y}_{c.T_c + H}, \dots, \mathbf{y}_{c.T_c + 1} \mid \mathbf{Y}_c, \mathbf{X}_c, \mathbf{A}_c, \boldsymbol\Sigma_c\right) \\
\times p\left( \mathbf{A}_c, \boldsymbol\Sigma_c \mid \mathbf{Y}_c, \mathbf{X}_c\right) d\left(\mathbf{A}_c, \boldsymbol\Sigma_c\right),\label{eq:pd}
\end{multline} where the forecast origin, \(T_c\), is a period for which
the last observation is available. Therefore, Bayesian forecasting
accounts for estimation uncertainty by integrating out the parameters
\(\mathbf{A}_c\) and \(\boldsymbol\Sigma_c\) from the predictive density
\citep[see][]{karlsson2013}. The estimation outcome, namely, the
posterior distribution of the parameters given data,
\(p\left( \mathbf{A}_c, \boldsymbol\Sigma_c \mid \mathbf{Y}_c, \mathbf{X}_c\right)\),
is used in the integration. Finally, the predictive density can easily
be formed using the conditional predictive density from equation
\eqref{eq:opapredictive}.

The numerical integration algorithm provides random draws from the
predictive density that can be used to provide forecast summaries of
interest, such as the mean and median forecast values, and the
predictive interval. These summaries are easy-to-communicate
characteristics of the predictive density.

\subsection{Sampling from Predictive Density}\label{ssec:sampling}

The density on the LHS of equation \eqref{eq:pd} is usually not of known
analytical form and, thus, the forecasting is performed by generating
random draws from the predictive density using numerical integration
techniques. The predictive density in \eqref{eq:pd} together with the
conditional density as defined in Section \ref{ssec:cpd} and the
one-period-ahead density in \eqref{eq:opapredictive} lead to the
following algorithm. It provides a sample from the predictive density
for country \(c\) using the sample of draws from the posterior
distribution of the country-specific parameters
\(\{\mathbf{A}_c^{(s)},\mathbf{\Sigma}_c^{(s)}\}_{s=1}^S\). Let \(T\)
denote both the sample size for the estimation of the model and the
forecast origin. We are interested in forecasting at horizons \(h\)
going from 1 to \(H\). The algorithm is as follows:

\begin{enumerate}
\item For each $s$ use the draw of parameters $\mathbf{A}_c^{(s)}$ and $\mathbf{\Sigma}_c^{(s)}$ to:
  \begin{enumerate}
  \item sample $\mathbf{y}_{T+1}^{(s)}$ from 
    $\mathcal{N}_N\left(\mathbf{A}_{c.1}^{(s)}\mathbf{y}_{c.T} + \dots + \mathbf{A}_{c.p}^{(s)} \mathbf{y}_{c.T-p+1} + \mathbf{A}_{c.d}^{(s)}\mathbf{d}_{c.T+1} , \boldsymbol\Sigma_c^{(s)}
\right)$
  \item for horizons $1<h\leq p$ sample $\mathbf{y}_{T+h}^{(s)}$ from 
    $$\mathcal{N}_N\left(\mathbf{A}_{c.1}^{(s)}\mathbf{y}_{c.T+h-1}^{(s)} + \dots + \mathbf{A}_{c.h-1}^{(s)}\mathbf{y}_{c.T+1}^{(s)} + \mathbf{A}_{c.h}^{(s)} \mathbf{y}_{c.T}+ \mathbf{A}_{c.p}^{(s)} \mathbf{y}_{c.T-p+h} +\dots + \mathbf{A}_{c.d}^{(s)}\mathbf{d}_{c.T+h} , \boldsymbol\Sigma_c^{(s)}
\right)$$
  \item for horizons $p < h \leq H$ sample $\mathbf{y}_{T+h}^{(s)}$ from 
    $$\mathcal{N}_N\left(\mathbf{A}_{c.1}^{(s)}\mathbf{y}_{c.T+h-1}^{(s)} + \dots + \mathbf{A}_{c.p}^{(s)}\mathbf{y}_{c.T-p+h}^{(s)} + \mathbf{A}_{c.d}^{(s)}\mathbf{d}_{c.T+h} , \boldsymbol\Sigma_c^{(s)}
\right)$$
  \end{enumerate}
\item repeat step 1. $S$ times for $s = 1,\dots, S$.
\item Return $\left\{\mathbf{y}_{T+1}^{(s)},\dots,\mathbf{y}_{T+H}^{(s)} \right\}_{s=1}^S$ as a sample drawn from the predictive density.
\end{enumerate}

Note that the algorithm is iterative in nature, that is, the draws of
forecasts at shorter horizons are used to construct the mean of the
density to sample those at further horizons. Also, note that the future
values of exogenous variables
\(\mathbf{d}_{c.T+1},\dots,\mathbf{d}_{c.T+H}\) are treated as given and
must be provided.

\subsection{Reporting Marginal Forecasts}

Institutional standards for forecast reporting often require the
presentation of forecasts for a subset of variables of interest despite
the modelling and forecasting frameworks including more variables.
Bayesian forecasting offers a simple and well-founded way to report
marginal forecasts for the subset of variables. To illustrate its
workings, consider a situation in which labour market forecasts are
reported but forecasted using also Gross Domestic Product. Divide the
vector \(\mathbf{y}_{c.t}\) into the GDP, \(gdp_{c.t}\), and labour
market variables, \(\mathbf{l}_{c.t}\), such that
\(\mathbf{y}_{c.t} = \begin{bmatrix} gdp_{c.t} & \mathbf{l}_{c.t}' \end{bmatrix}'\).
Then the joint predictive density can be expressed considering the
corresponding block of the mean and covariance matrix: \begin{align}
p\left( \begin{bmatrix} gdp_{c.t+1} \\ \mathbf{l}_{c.t+1} \end{bmatrix} \mid 
 \begin{bmatrix} gdp_{c.t} \\ \mathbf{l}_{c.t} \end{bmatrix}, \dots,  \begin{bmatrix} gdp_{c.t-p+1} \\ \mathbf{l}_{c.t-p+1} \end{bmatrix},
\mathbf{A}_c, \boldsymbol\Sigma_c\right) = \mathcal{N}_N\left( \begin{bmatrix}m_{c.t+1}^{(gdp)} \\ \mathbf{m}_{c.t+1}^{(l)} \end{bmatrix}, \begin{bmatrix} \sigma_c^{2(gdp)} & \boldsymbol\Sigma_{c}^{(l.gdp)\prime} \\ \boldsymbol\Sigma_{c}^{(l.gdp)} &  \boldsymbol\Sigma_{c}^{(l)} \end{bmatrix} \right)\label{eq:predictive_factored}
\end{align} where \(\mathbf{m}_{c.t+1}^{(l)}\) and \(m_{c.t+1}^{(gdp)}\)
are appropriately factorised elements of the predictive density mean,
and \(\boldsymbol\Sigma_{c}^{(l)}\),
\(\boldsymbol\Sigma_{c}^{(l.gdp)}\), and \(\sigma_c^{2(gdp)}\) are those
for the covariance matrix \(\boldsymbol\Sigma_c\).

Reporting marginal forecasts is based on two results. Firstly, the
marginal predictive density for the labour market outcomes when the
joint predictive density is as in equation
\eqref{eq:predictive_factored} is simply given by \begin{align}
p\left( \mathbf{l}_{c.t+1} \mid 
 \begin{bmatrix} gdp_{c.t} \\ \mathbf{l}_{c.t} \end{bmatrix}, \dots,  \begin{bmatrix} gdp_{c.t-p+1} \\ \mathbf{l}_{c.t-p+1} \end{bmatrix},
\mathbf{A}_c, \boldsymbol\Sigma_c\right) = \mathcal{N}_{N-1}\left(  \mathbf{m}_{c.t+1}^{(l)}, \boldsymbol\Sigma_{c}^{(l)} \right)\label{eq:predictive_marginal}
\end{align} The density in \eqref{eq:predictive_marginal} is the
marginal predictive density of labour market variables,
\(\mathbf{l}_{c.t+1}\). However it still depends on the past GDP values.

For one-period-ahead forecasting the task is finished because the past
values of GDP are sample data. Labour market variables forecasts at more
distant horizons will, however, depend on GDP forecasts. This is
addressed by marginalisation of the predictive density over these GDP
forecasts. This marginalisation is formally performed by integration of
the joint density of GDP and labour market variables forecasts over the
former values. To illustrate this integration and for simplicity of
exposition without loosing the generality, consider two-period-ahead
forecasts, \(h=2\), performed by a model with one autoregressive lag,
\(p=1\). Then the marginal density forecast for the labour variables at
horizon one and two is given by: \begin{align}
p\left( \mathbf{l}_{c.t+2}, \mathbf{l}_{c.t+1} \mid 
\mathbf{A}_c, \boldsymbol\Sigma_c\right) &= 
\int p\left( \mathbf{l}_{c.t+2}, \mathbf{l}_{c.t+1}, gdp_{c.t+1} \mid
\mathbf{A}_c, \boldsymbol\Sigma_c\right) dgdp_{c.t+1}\label{eq:prefl1}\\
&= 
\int p\left( \mathbf{l}_{c.t+2}\mid \mathbf{l}_{c.t+1}, gdp_{c.t+1},
\mathbf{A}_c, \boldsymbol\Sigma_c\right) p\left(\mathbf{l}_{c.t+1}, gdp_{c.t+1} \mid \mathbf{l}_{c.t}, gdp_{c.t},
\mathbf{A}_c, \boldsymbol\Sigma_c\right) dgdp_{c.t+1}.\label{eq:prefl2}
\end{align} It is obtained by integrating out one-period-ahead forecast
of GDP from the joint predictive density. This joint predictive density
is constructed in \eqref{eq:prefl2} by factorising it into the marginal
two-period ahead predictive density for \(\mathbf{l}_{c.t+2}\) given by
\(p\left( \mathbf{l}_{c.t+2}\mid \mathbf{l}_{c.t+1}, gdp_{c.t+1},
\mathbf{A}_c, \boldsymbol\Sigma_c\right)\) as defined in
\eqref{eq:predictive_marginal}, and the joint one-period-ahead density
\(p\left(\mathbf{l}_{c.t+1}, gdp_{c.t+1} \mid \mathbf{l}_{c.t}, gdp_{c.t}, \mathbf{A}_c, \boldsymbol\Sigma_c\right)\)
as defined in \eqref{eq:predictive_factored}.

The simplicity of this solution is that the integral is solved
numerically automatically in the sampling procedure for Bayesian
forecasting presented in Section \ref{ssec:sampling}. It suffices to
report the summary statistics of labour market outcomes forecasts,
without the DGP forecasts, to obtain the demanded effect. The
interpretation of such forecast is that labour market forecasts are
averaged over all \(S\) paths of future GDP simulated from its
predictive density.

\subsection{Conditional Projections}

\noindent The International Monetary Fund's GDP projections for all 189
countries are borrowed into our VAR setup. At the same time, including
GDP as one of the endogenous variables in the Vector Autoregression is a
necessary feature of a reliable forecasting model. In this context,
following the argument by \cite{Sims1980}, we include GDP as if it was
an endogenous variable in the Vector Autoregression for modelling labour
market outcomes. We proceed this way in order to estimate the model and
form the joint predictive density. This enables us to reliably predict
labour market outcomes based on the future trajectories of GDP.

Such conditional forecasting is an indispensable feature of Bayesian
forecasting \citep[see][]{Doan1984,waggoner_conditional_1999}. The
one-period-ahead predictive density in expression
\eqref{eq:opapredictive} as factorised in equation
\eqref{eq:predictive_factored} is the basis for our implementation of
such forecasts. Then, forecasting of the labour market outcomes,
\(\mathbf{l}_{c.t+1}\), given the contemporaneous projection of GDP,
\(gdp_{c.t}\), is performed using the conditional predictive density:
\begin{multline}
p\left(  \mathbf{l}_{c.t+1} \mid gdp_{c.t+1}, \mathbf{y}_{c.t}, \dots, \mathbf{y}_{c.t-p+1}, \mathbf{A}_c, \boldsymbol\Sigma_c\right) =\\ \mathcal{N}_{N-1}\left( \mathbf{m}_{c.t+1}^{(l)} + \boldsymbol\Sigma_{c}^{(l.gdp)} \sigma_c^{-2(gdp)}\left( gdp_{c.t+1} - m_{c.t+1}^{(gdp)}\right),  \boldsymbol\Sigma_{c}^{(l)} - \boldsymbol\Sigma_{c}^{(l.gdp)} \sigma_c^{-2(gdp)} \boldsymbol\Sigma_{c}^{(l.gdp)\prime}  \right)\label{eq:condfore}
\end{multline} In conditional forecasting as in equation
\eqref{eq:condfore} GDP is never predicted. Instead, its projections are
used to forecast labour market outcomes.

\subsection{Restricted Forecasts}

The approach to forecasting described in the current paper implements
optimal choices based on the best Bayesian forecasting practices, thus
leading to reliable predictions. This means we forecast labour market
rates without applying any transformations such as taking the logarithm
or differencing. The rates must be within the interval \([0,100]\) and
the forecasts must respect this constraint. This is ensured following
the approach by \cite{waggoner_conditional_1999}, which transforms the
one-period-ahead predictive densities specified in equations
\eqref{eq:opapredictive} and \eqref{eq:condfore} into truncated
multivariate normal distributions between \([0,100]\). The numerical
implementation is based on the function \texttt{mvrandn()} from the
package \pkg{TruncatedNormal} by \cite{TruncatedNormal} implementing the
constrained sampler by \cite{Botev}. This forecasting method can be
implemented without further adjustments to the estimation algorithm.

\subsection{Pseudo-out-of-sample Forecasting}

We implement pseudo-out-of-sample forecasting exercise to check the
external validity of the models. This task is facilitated by using the
available data to verify the forecast performance. The sample is split
into training and forecast evaluation samples with the cut-off point
\(T_{f_1} < T_c\), where the observations up to time \(T_{f_1}\) are
used for the estimation of the model. Then, we forecast at the
application-driven horizons of interest, say one and two periods ahead,
\(T_{f_1} + 1\) and \(T_{f_1} + 2\) respectively. The maximum value of
the forecasting horizon is denoted by \(H\) and in this example is equal
to \(H=2\). The observations at periods \(T_{f_1} + 1\) and
\(T_{f_1} + 2\) are used to verify prediction precision according to
selected measures. This estimation and forecasting exercise is repeated
iteratively using an expanding window of the training sample, that is,
the cut-off point \(T_{f_1}\) is moved forward by one period at a time.
For instance, in the second iteration of the exercise the cut-off point
for the estimation sample is set to \(T_{f_2} = T_{f_1} + 1\).
Altogether, the exercise is repeated \(F = T_c - T_{f_1} - H + 1\)
times. At each iteration, the models are re-estimated, and forecasts are
saved. This procedure uses parallel computations using \pkg{OpenMP} by
\cite{dagum1998openmp} for each of the iterations.

\subsection{Forecast Performance Measures}

The pseudo-out-of-sample forecasting exercise is complemented by
calculating the forecast performance measures. Let the forecasting
horizon of interest be denoted by \(h\) and the observations used to
verify forecast precision for a country \(c\) by
\(\mathbf{y}_{c.T_{f_1}+h}, \dots, \mathbf{y}_{c.T_{f_{F}}+h}\).
Consider point forecasting with the predictions given by the mean of the
predictive density denoted by
\(\bar{\mathbf{y}}_{c.T_{f_i}+h} = S^{-1}\sum_{s=1}^{S}\mathbf{y}_{c.T_{f_i}+h}^{(s_)}\)
for \(i = 1,\dots,F\). Then the root-mean-squared-forecast error
(\(RMSFE\)) for the \(n^{\text{th}}\) variable for forecasting horizon
\(h\) is defined as \begin{align}
RMSFE_n = \sqrt{\frac{1}{F}\sum_{i=1}^F \left( y_{c.T_{f_i}+h.n} - \bar{y}_{c.T_{f_i}+h.n}\right)^2}.
\end{align} The RMSFE can be averaged across all variables using
\(RMSFE = \sqrt{N^{-1}\sum_{n=1}^N RMSFE_n^2}\). As an alternative
measure of point forecast performance, the mean-absolute-forecast error
(\(MAFE\)) for the \(n^{\text{th}}\) variable for horizon \(h\) is
defined as \begin{align}
MAFE_n = \frac{1}{F}\sum_{i=1}^F \left| y_{c.T_{f_i}+h.n} - \bar{y}_{c.T_{f_i}+h.n}\right|,
\end{align} and is aggregated to a joint measure using
\(MAFE = N^{-1}\sum_{n=1}^N MAFE_n\). Both of the measures, in their
single-variable and averaged versions, can be averaged further across
countries by applying similar averaging formulae. For both, \(RMSFE\)
and \(MAFE\), the lower the value, the better the forecast performance.

As a performance measure of density forecasts consider the predictive
log-score (PLS) of a model defined by \cite{geweke_comparing_2010} as
\begin{align}
PLS = \frac{1}{F}\sum_{i=1}^F \log \hat{p}\left(\mathbf{y}_{c.T_{f_i}+h} \mid \mathbf{y}_{c.T_{f_i}}, \dots, \mathbf{y}_{c.T_{f_i}-p+1}\right),
\end{align} where the predictive density ordinates
\(\hat{p}\left(\mathbf{y}_{c.T_{f_i}+h} \mid \mathbf{Y}_{c.T_{f_i}}, \mathbf{X}_{c.T_{f_i}}\right)\),
following \cite{gelfand1990sampling}, are estimated through numerical
integration using the sample from the posterior distribution of the
parameters \begin{align}
\hat{p}\left(\mathbf{y}_{c.T_{f_i}+h} \mid \mathbf{y}_{c.T_{f_i}}, \dots, \mathbf{y}_{c.T_{f_i}-p+1}\right) = \frac{1}{S}\sum_{s=1}^S p\left(\mathbf{y}_{c.T_{f_i}+h} \mid \mathbf{y}_{c.T_{f_i}}, \dots, \mathbf{y}_{c.T_{f_i}-p+1}, \mathbf{A}_c^{(s)}, \boldsymbol\Sigma_c^{(s)}\right),
\end{align} where the conditional densities on the right-hand side are
defined in expression \eqref{eq:opapredictive}. \(PLS\) can be reported
for individual variables using their marginal predictive densities or
for the whole system as above. It is also subject to aggregation over
countries. The higher the \(PLS\) value, the better the forecast
performance.

\section{Using the R Package bpvars}\label{sec:flmo}

The methods proposed in this paper are implemented in the \proglang{R}
Package \pkg{bpvars} by \cite{bpvars}. The package offers a simple
workflow for the model specification, estimation, forecasting, and their
summaries and visualisations. This section first presents the basic
workflow of the package and then focuses on its particular steps. This
is followed by the presentation of the pseudo-out-of-sample forecasting
exercise.

\subsection{The Basic Workflow}

Load the package to the \proglang{R} environment by executing the
following line:

\begin{CodeChunk}
\begin{CodeInput}
R> library(bpvars)
\end{CodeInput}
\end{CodeChunk}

The dynamic panel data to be used for estimation must be formatted as a
list containing in its elements \texttt{ts} matrices with
country-specific time series with \(T_c\) rows and \(N\) columns. The
package provides appropriately formatted sample data for a system
containing gross domestic product, unemployment rate, employment rate,
and labour force participation rate in object
\texttt{ilo\_dynamic\_panel}. It is automatically loaded upon package
loading. Display several first observations for Australia by running

\begin{CodeChunk}
\begin{CodeInput}
R> head(ilo_dynamic_panel$AUS)
\end{CodeInput}
\begin{CodeOutput}
Time Series:
Start = 1991 
End = 1996 
Frequency = 1 
          gdp    UR      EPR LFPR
1991 27.16421  9.59 57.21842 63.3
1992 27.16899 10.70 56.23992 63.0
1993 27.20790 10.90 55.78678 62.6
1994 27.24683  9.72 56.94337 63.1
1995 27.28571  8.47 58.28255 63.7
1996 27.32314  8.51 58.23320 63.6
\end{CodeOutput}
\end{CodeChunk}

Specify the model by running a simple function that specifies the model
and all the required values, such as the number of lags, data matrices
for individual countries, prior distribution hyper-parameters, starting
values for the Gibbs sampler, and some characteristics for specific
steps of the estimation algorithm. For instance, the code below relies
on the default model setup and will specify a model with one lag, and
the Minnesota prior with the prior mean for the autoregressive
parameters reflecting the unit-root non-stationarity of the variables
and save the model specification in object \texttt{spec} that is of
class \texttt{BVARPANEL}.

\begin{CodeChunk}
\begin{CodeInput}
R> spec = specify_bvarPANEL$new(ilo_dynamic_panel)
\end{CodeInput}
\end{CodeChunk}

This object is then provided as the first argument of the function
\texttt{estimate()} that runs the initial \texttt{1000} iterations of
the Gibbs sampler described in Appendix \ref{sec:gibbs} in the burn-in
phase:

\begin{CodeChunk}
\begin{CodeInput}
R> burn = estimate(spec, S = 1000)
\end{CodeInput}
\begin{CodeOutput}
**************************************************|
bpvars: Forecasting with Bayesian Panel VARs      |
**************************************************|
 Progress of the MCMC simulation for 1000 draws
    Every draw is saved via MCMC thinning
 Press Esc to interrupt the computations
**************************************************|
\end{CodeOutput}
\end{CodeChunk}

The code above reads the first argument and applies appropriate
algorithms to estimate the model. More precisely, the fact that the
first argument, that is the object \texttt{spec}, is of class
\texttt{BVARPANEL} triggers the execution of method
\texttt{estimate.BVARPANEL()} that reads the starting values and runs
the Gibbs sampler. The object \texttt{burn} is of class
\texttt{PosteriorBVARPANEL}.

As the first run achieves convergence, the second execution of the
function \texttt{estimate()} will provide the final \texttt{1000} draws
from the posterior distribution. To facilitate this provide object
\texttt{burn} as the first argument to the \texttt{estimate()} function:

\begin{CodeChunk}
\begin{CodeInput}
R> post = estimate(burn, S = 1000)
\end{CodeInput}
\begin{CodeOutput}
**************************************************|
bpvars: Forecasting with Bayesian Panel VARs      |
**************************************************|
 Progress of the MCMC simulation for 1000 draws
    Every draw is saved via MCMC thinning
 Press Esc to interrupt the computations
**************************************************|
\end{CodeOutput}
\end{CodeChunk}

This execution of the function \texttt{estimate()} is determined by the
class of the object \texttt{burn} and triggers the estimation algorithm
using method \texttt{estimate.PosteriorBVARPANEL()}. This method reads
the last draw from \texttt{burn}, sets it as the starting value, and
continues the Gibbs sampler to obtain the final \texttt{1000} draws from
the target posterior distribution. The object \texttt{post} is also of
class \texttt{PosteriorBVARPANEL}.

Investigate the estimates of the autoregressive parameters for the
second variable, that is unemployment rate \texttt{UR}, for Australia
denoted by the ISO country code \texttt{AUS} by applying the method
\texttt{summary()} to the estimation outcome, saving it in object
\texttt{post\_summ}, and extracting the list element called after the
ISO code of the country and then extracting element \texttt{A} standing
for the autoregressive matrix:

\begin{CodeChunk}
\begin{CodeInput}
R> post_summ = summary(post)
R> post_summ$AUS$A$equation2
\end{CodeInput}
\begin{CodeOutput}
                mean         sd 5% quantile 95% quantile
lag1_var1 -1.1505725  0.6156524  -2.1181635   -0.2060646
lag1_var2 -0.3078682  0.9153885  -1.7920723    1.1529590
lag1_var3 -1.8618742  1.4507439  -4.2367150    0.4577542
lag1_var4  2.0033177  1.3937978  -0.2688551    4.3137686
const     23.2569027 14.9260972  -1.6330710   47.0733018
\end{CodeOutput}
\end{CodeChunk}

The displayed characteristics of the posterior distribution of the
parameters in this equation include the mean, standard deviation, as
well as the 5\% and 95\% quantiles.

Forecast the labour market outcomes and the GDP three years ahead by
running the function \texttt{forecast()}, setting its argument
\texttt{horizon} to \texttt{3}, and saving its output in an object
\texttt{fore}:

\begin{CodeChunk}
\begin{CodeInput}
R> fore = forecast(post, horizon = 3)
\end{CodeInput}
\end{CodeChunk}

This function uses the draws from the posterior distribution in object
\texttt{post} and applies Bayesian forecasting procedure described in
Section \ref{sec:fore}. The object \texttt{fore} is of class
\texttt{ForecastBVARPANEL} and can be used to summarise and visualise
the forecasting results. Apply the \texttt{summary()} function choosing
the country-specific forecasts by setting the argument \texttt{which\_c}
to the ISO code of the country, and display the forecasts for the second
variable, that is, the unemployment rate. The table can be interpreted
as reporting the characteristics of the marginal predictive density of
the unemployment rate.

\begin{CodeChunk}
\begin{CodeInput}
R> fore_sum = summary(fore, which_c = "AUS")
\end{CodeInput}
\begin{CodeOutput}
 **************************************************|
 bsvars: Bayesian Structural Vector Autoregressions|
 **************************************************|
   Posterior summary of forecasts                  |
 **************************************************|
\end{CodeOutput}
\begin{CodeInput}
R> fore_sum$variable2 
\end{CodeInput}
\begin{CodeOutput}
      mean        sd 5% quantile 95% quantile
1 3.984685 0.6347587    2.931214     4.975218
2 4.055137 0.8547802    2.638279     5.471585
3 4.102732 1.0240078    2.414115     5.707222
\end{CodeOutput}
\end{CodeChunk}

Visualise the forecasts for Australia in Figure~2 by applying method
\texttt{plot()} and setting its argument \texttt{which\_c} to the ISO
code of the country:

\begin{CodeChunk}
\begin{CodeInput}
R> plot(fore, which_c = "AUS") 
\end{CodeInput}
\begin{figure}

{\centering \includegraphics{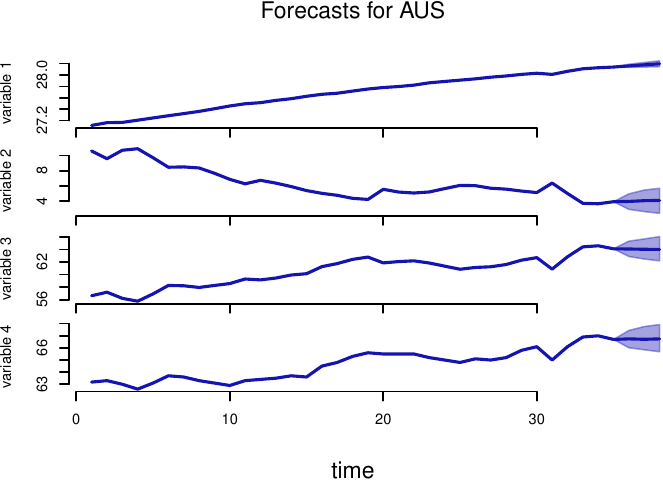} 

}

\caption[Three years ahead forecasts for Australia]{Three years ahead forecasts for Australia}\label{fig:fore_plot}
\end{figure}
\end{CodeChunk}

Compute the forecast error variance decomposition by running the
function \texttt{compute\_variance\_decompositions()} and saving its
output in object \texttt{fevd}. The forecast horizon is determined by
argument \texttt{horizon} set to \texttt{3} and the choice of the
country is made using the argument \texttt{which\_c}. The plot is given
in Figure~3.

\begin{CodeChunk}
\begin{CodeInput}
R> fevd = compute_variance_decompositions(post, horizon = 3)
R> plot(fevd, which_c = "AUS")
\end{CodeInput}
\begin{figure}

{\centering \includegraphics{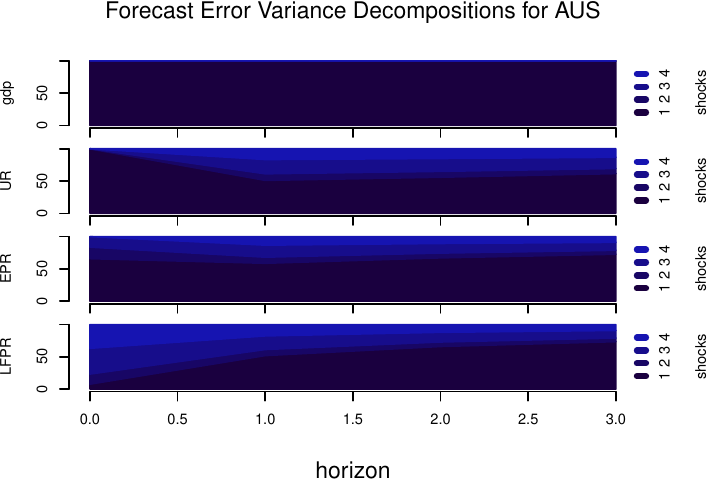} 

}

\caption[Three years ahead forecast error variance decompositions for Australia]{Three years ahead forecast error variance decompositions for Australia}\label{fig:fevd}
\end{figure}
\end{CodeChunk}

This basic workflow can be coded using the pipe operator
\texttt{\textbar{}\textgreater{}} that provides the object obtained by
the code preceding it as the first argument of the function following
it. We propose to separate the estimation part from the forecasting one.
The workflow below first provides the data to the function specifying
the model, which is provided to the \texttt{estimate()} function to run
the initial 1000 draws of the burn-in stage, which in turn is provided
to the \texttt{estimate()} function to obtain the final 1000 iterations
of the Gibbs sampler. The single output of this part is saved in object
\texttt{post}:

\begin{CodeChunk}
\begin{CodeInput}
R> ilo_dynamic_panel |> 
+   specify_bvarPANEL$new() |> 
+   estimate(S = 1000, show_progress = FALSE) |> 
+   estimate(S = 1000, show_progress = FALSE) -> post
\end{CodeInput}
\end{CodeChunk}

Having estimated the model, forecasting is executed and its outcome can
be plotted using:

\begin{CodeChunk}
\begin{CodeInput}
R> post |> 
+   forecast(horizon = 3)  |> 
+   plot(which_c = "AUS", main = "Forecasts for Australia")
\end{CodeInput}
\end{CodeChunk}

Similarly, the forecast error variance decompositions can computed and
plotted:

\begin{CodeChunk}
\begin{CodeInput}
R> post |> 
+   compute_variance_decompositions(horizon = 3) |> 
+   plot(which_c = "AUS", main = "Forecast Error Variance Decompositions for Australia")
\end{CodeInput}
\end{CodeChunk}

In what follows, we focus on particular elements of the workflow
described above. They are presented expanding the estimation and
forecasting to a more realistic scenario using a model specification
with exogenous variables, and with conditional forecasts given the
projections of GDP.

\subsection{Install the Package}\label{ssec:install}

The package is installed once prior to working with it from the CRAN
repository by running the command:

\begin{CodeChunk}
\begin{CodeInput}
R> install.packages("bpvars")
\end{CodeInput}
\end{CodeChunk}

The correct functioning of the \pkg{bpvars} requires the installation of
the \proglang{R} package \pkg{bsvars} and other dependencies, which will
proceed automatically during the installation of the \pkg{bpvars}
package. Before every use of the package load it to the memory by
running the command:

\begin{CodeChunk}
\begin{CodeInput}
R> library(bpvars)
\end{CodeInput}
\end{CodeChunk}

\subsection{Get Familiar with Data}\label{ssec:data}

In order to implement the model with dummy variables, the package
provides data objects formatted as required by the package functions.
The dynamic panel data set contains annual observations from 1991 to
2024 for the logarithm of gross domestic product (\(gdp\)), unemployment
rate (\(UR\)), employment rate (\(EPR\)), labour force participation
rate (\(LFPR\)) for 189 countries are downloaded from \cite{ILO} using
the \proglang{R} Package \pkg{Rilostat} by \cite{Rilostat}. The GDP data
are taken from the IMF database. Figure 4 plots all time series
contained in the data object \texttt{ilo\_dynamic\_panel} and highlights
the series for Australia in each of the plots.

\begin{CodeChunk}
\begin{figure}

{\centering \includegraphics{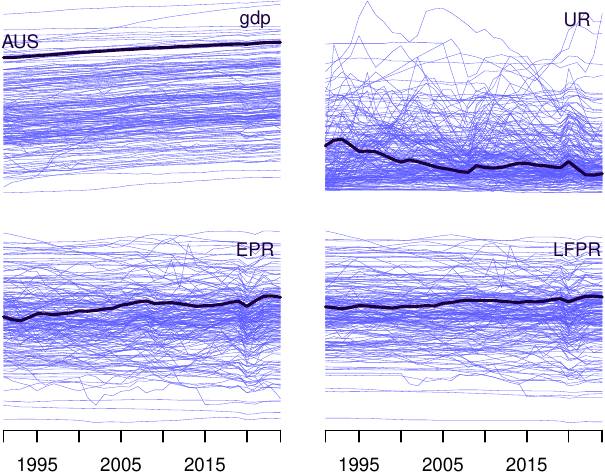} 

}

\caption[Dynamic panel data for a four variable system]{Dynamic panel data for a four variable system}\label{fig:fig-dataplot}
\end{figure}
\end{CodeChunk}

Inspect the dummy variables for year 2008 of the global financial
crisis, and years 2020 and 2021 of the COVID pandemic:

\begin{CodeChunk}
\begin{CodeInput}
R> tail(ilo_exogenous_variables$POL)
\end{CodeInput}
\begin{CodeOutput}
Time Series:
Start = 2019 
End = 2024 
Frequency = 1 
     2008 2020 2021
2019    0    0    0
2020    0    1    0
2021    0    0    1
2022    0    0    0
2023    0    0    0
2024    0    0    0
\end{CodeOutput}
\end{CodeChunk}

The forecasted values of these dummy variables are necessary for
forecasting:

\begin{CodeChunk}
\begin{CodeInput}
R> ilo_exogenous_forecasts$POL
\end{CodeInput}
\begin{CodeOutput}
Time Series:
Start = 2025 
End = 2027 
Frequency = 1 
     2008 2020 2021
2025    0    0    0
2026    0    0    0
2027    0    0    0
\end{CodeOutput}
\end{CodeChunk}

Each of these data objects features documentation that can be accessed
by the \texttt{?} operator, for example:

\begin{CodeChunk}
\begin{CodeInput}
R> ?ilo_exogenous_forecasts
\end{CodeInput}
\end{CodeChunk}

\subsection{Specify the Model}

The function specifying the model presented in Section \ref{sec:hpvar}
provides ample possibilities of modifying the model. In order to
implement the model with exogenous variables, one autoregressive lag,
for the labour market variables and GDP, one needs to execute:

\begin{CodeChunk}
\begin{CodeInput}
R> spec = specify_bvarPANEL$new( 
+   data = ilo_dynamic_panel, 
+   p = 1,  
+   exogenous = ilo_exogenous_variables,
+   stationary = c(FALSE, FALSE, FALSE, FALSE)
+ )
\end{CodeInput}
\end{CodeChunk}

The \texttt{spec} object is of class \texttt{BVARPANEL} and includes the
model specification. The arguments of the function
\texttt{specify\_bvarPANEL\$new()} provide the possibility of
implementing the basic setup of the model. The argument \texttt{data}
reads appropriately formatted data, the argument \texttt{p} specifies
the number of autoregressive lags, and the argument \texttt{exogenous}
reads the exogenous variables. Finally, the argument \texttt{stationary}
specifies whether the variables are stationary, which subsequently sets
the Minnesota prior mean for the global autoregressive matrix
\(\mathbf{A}\). A value \texttt{FALSE} sets the prior mean to the random
walk process for the particular variable, whereas \texttt{TRUE} sets it
to white noise process.

Further investigation or customisation of the model specification can be
made by displaying or altering the elements of the \texttt{spec} object.
For instance, to display the matrix \(\underline{\mathbf{M}}\) from
equation \eqref{eq:priorA} showing the prior mean of the global
autoregressive matrix, execute:

\begin{CodeChunk}
\begin{CodeInput}
R> spec$prior$M
\end{CodeInput}
\begin{CodeOutput}
     [,1] [,2] [,3] [,4]
[1,]    1    0    0    0
[2,]    0    1    0    0
[3,]    0    0    1    0
[4,]    0    0    0    1
[5,]    0    0    0    0
[6,]    0    0    0    0
[7,]    0    0    0    0
[8,]    0    0    0    0
\end{CodeOutput}
\end{CodeChunk}

This \(8\times 4\) matrix includes the autoregressive matrix in the top
\(4\times 4\), the prior mean for the constant term vector in the
\(5^{\text{th}}\) row, and the prior means for the global slopes on the
dummies in the last three rows. The ones on the main diagonal mean that
the overall global persistence hyper-parameter \(m\) is estimated. This
global autoregressive matrix prior mean can be changed to the pooled
estimator by \cite{zellner_hong} by running
\texttt{spec\$set\_global2pooled()} similarly and the model with the
prior by \cite{jarocinski_responses_2010} can be set by running
\texttt{spec\$set\_to\_Jarocinski()}.

Any element of the model specification can be altered but this should be
done with care, making certain that the class of the provided object
aligns with the class of the element being altered, and that the
provided value is appropriately solicited. This includes all the fixed
prior hyper-parameters from Section \ref{ssec:fphp} that are coded as
elements of list in object \texttt{spec\$prior}. For instance, consider
the prior mean \(\underline{\lambda}\) of the shape parameter \(\nu\)
from equation \eqref{eq:priorNU} whose default value can be accessed by:

\begin{CodeChunk}
\begin{CodeInput}
R> spec$prior$lambda 
\end{CodeInput}
\begin{CodeOutput}
[1] 72
\end{CodeOutput}
\end{CodeChunk}

Consider setting the median of this exponential prior distribution to
thirty, which can be implemented and verified by running:

\begin{CodeChunk}
\begin{CodeInput}
R> spec$prior$lambda = 43.29
R> qexp(0.5, 1/spec$prior$lambda) 
\end{CodeInput}
\begin{CodeOutput}
[1] 30.00634
\end{CodeOutput}
\end{CodeChunk}

Finally, the documentation for the model specification can be accessed
by running \texttt{?specify\_bvarPANEL} with details on its particular
parts, such as the function generating the model specification
\texttt{specify\_bvarPANEL\$new()}, available following the links in the
documentation just displayed.

\subsection{Specify the Model with Country Grouping}

The specification of the model with country grouping is similar to the
one presented in Section \ref{ssec:specCG} and uses facility
\texttt{specify\_bvarGroupPANEL}. To specify a model with fixed group
allocations the argument \texttt{group\_allocation} must be provided as
a vector of length \(C\) with integers denoting the groups \(g\). The
package provides several such vectors that group countries by their
geographical location with various levels of aggregation, see vectors
\texttt{country\_grouping\_region},
\texttt{country\_grouping\_subregionbroad}, and
\texttt{country\_grouping\_subregiondetailed}, and by their income level
implmented in \texttt{country\_grouping\_incomegroup}. The example below
uses the vector \texttt{country\_grouping\_region}.

\begin{CodeChunk}
\begin{CodeInput}
R> spec_fg = specify_bvarGroupPANEL$new(
+   data = ilo_dynamic_panel, 
+   p = 1,  
+   exogenous = ilo_exogenous_variables,
+   stationary = c(FALSE, FALSE, FALSE, FALSE),
+   group_allocation = country_grouping_region
+ )
\end{CodeInput}
\end{CodeChunk}

Alternatively, the group allocation can be estimated. Such a model
requires the specification of the number of groups \(G\) using the
argument \texttt{G} as in the example below where it is set to value
\texttt{2}.

\begin{CodeChunk}
\begin{CodeInput}
R> spec_eg = specify_bvarGroupPANEL$new(
+   data = ilo_dynamic_panel, 
+   p = 1,  
+   exogenous = ilo_exogenous_variables,
+   stationary = c(FALSE, FALSE, FALSE, FALSE),
+   G = 2
+ )
\end{CodeInput}
\end{CodeChunk}

The \texttt{spec} objects created in the listings above can be used in
the workflows that follow. Consult the package documentation on the
model specification using \texttt{?specify\_bvarGroupPANEL} or on the
provided country groupings using, for instance,
\texttt{?country\_grouping\_region}.

\subsection{Estimate the Model}

The function \texttt{estimate()} runs the Gibbs sampler and is
implemented as two methods, \texttt{estimate.BVARPANEL()} and
\texttt{estimate.PosteriorBVARPANEL()}. It starts the estimation
respectively at the starting values from the model specification object
of class \texttt{BVARPANEL} or from the last draw of the previous run
provided in an object of class \texttt{PosteriorBVARPANEL}. The Gibbs
sampler presented in Appendix \ref{sec:gibbs} performs iterations
sampling random draws from the full conditional posterior distributions
of the parameters of the model, namely, the country-specific parameter
matrices \(\mathbf{A}_c\) and \(\mathbf{\Sigma}_c\), the global
parameter matrices \(\mathbf{A}\) and \(\mathbf{\Sigma}\), the
hierarchical prior estimated parameters \(\mathbf{V}\), \(\nu\), \(m\),
\(w\), and \(s\), and missing observations if required.

The \texttt{estimate()} function uses argument \texttt{S} to set the
number of Gibbs sampler iterations, and argument \texttt{thin} to set
the thinning parameter. Thinning is a procedure to reduce the
autocorrelation in the posterior draws, rendering the posterior
estimates more efficient. Alternatively, this function is used to manage
the computer memory used for estimation. It is obtained by returning
every \texttt{thin} draw in the final sample. This procedure reduces the
number of draws returned by the \texttt{estimate()} function to
\texttt{S/thin}. The last argument of this function is
\texttt{show\_progress} that gives users the choice of whether or not to
display the progress bar. Finally, the documentation can be accessed by
running \texttt{?estimate.BVARPANEL} or
\texttt{?estimate.PosteriorBVARPANEL}.

The model is estimated in two stages, the burn-in and the final one both
run for \texttt{1000} iterations:

\begin{CodeChunk}
\begin{CodeInput}
R> burn = estimate(spec, S = 1000, show_progress = FALSE)    
R> post = estimate(burn, S = 1000, show_progress = FALSE) 
\end{CodeInput}
\end{CodeChunk}

and the posterior estimates can be computed using the \texttt{summary()}
function, for instance, executing:

\begin{CodeChunk}
\begin{CodeInput}
R> post_summ = summary(post)   
R> post_summ$POL$Sigma$equation4 
\end{CodeInput}
\begin{CodeOutput}
                   mean          sd  5% quantile 95% quantile
Sigma[4,1] -0.002348656 0.001332732 -0.004648954 -0.000350491
Sigma[4,2]  0.130546735 0.102237333 -0.030313425  0.302997134
Sigma[4,3]  0.051495879 0.056796938 -0.034724591  0.151172363
Sigma[4,4]  0.150744594 0.035734830  0.100835300  0.213684899
\end{CodeOutput}
\end{CodeChunk}

reports the estimates of the fourth row of Poland's error term
covariance matrix.

Based on authors' experience of working with the package, the data set
\texttt{ilo\_dynamic\_panel}, and a model with exogenous variables
\texttt{ilo\_exogenous\_variables} and \(p=1\) lag, using at least
\texttt{S\ =\ 1000} iterations is recommended for both burn-in and final
estimation. More iterations for both of the estimation stages might be
required for data sets with a larger number of variables or a model with
more lags.

\subsection{Forecast Labour Market Outcomes}

The function \texttt{forecast()} implements method
\texttt{forecast.PosteriorBVARPANEL()} that provides draws from the
predictive density as discussed in Section \ref{sec:fore}. This
forecasting routine performs forecasting for all countries and
facilitates conditional forecasting given the future trajectories of
some variables as well as forecasting for models with exogenous
variables.

The example below presents the full extent of the forecasting
capabilities of the \texttt{forecast()} function. Its first argument is
the object \texttt{post} of class \texttt{PosteriorBVARPANEL} that
contains the posterior draws from the model's posterior distribution.
The remaining arguments must be aligned in terms of the forecasting
horizon. The code below sets the argument \texttt{horizon} to value
three, which requires the lists provided to the last argument, namely
\texttt{exogenous\_forecast}, to contain matrices with three rows.
Appropriately constructed objects provided as arguments are described in
Section \ref{ssec:data}. The outcome of the forecasting procedure is
stored in the object \texttt{fore} of class \texttt{ForecastsPANEL}.

\begin{CodeChunk}
\begin{CodeInput}
R> fore = forecast(
+   post,   
+   horizon = 3,
+   exogenous_forecast = ilo_exogenous_forecasts 
+ ) 
\end{CodeInput}
\end{CodeChunk}

The format of these arguments and the output is explained in the package
documentation available by running
\texttt{?forecast.PosteriorBVARPANEL}.

\subsection{Report the Forecasts}

Having obtained the draws from the predictive density saved in object
\texttt{fore}, the user can report the forecasts for the variable of
interest. The function \texttt{summary()} provides the summary
statistics of the forecasts such as the mean, standard deviation, and
the 5\% and 95\% quantiles for Poland. The choice of country is made by
setting argument \texttt{which\_c} to value \texttt{"POL"}.

\begin{CodeChunk}
\begin{CodeInput}
R> fore_sum = summary(fore, which_c = "POL") 
\end{CodeInput}
\begin{CodeOutput}
 **************************************************|
 bsvars: Bayesian Structural Vector Autoregressions|
 **************************************************|
   Posterior summary of forecasts                  |
 **************************************************|
\end{CodeOutput}
\begin{CodeInput}
R> fore_sum$variable3  
\end{CodeInput}
\begin{CodeOutput}
      mean        sd 5% quantile 95% quantile
1 57.47651 0.9200653    55.92822     59.05520
2 57.95958 1.3949942    55.65229     60.22082
3 58.47155 1.7628337    55.67826     61.33150
\end{CodeOutput}
\end{CodeChunk}

Finally, the forecasts can be visualised using the \texttt{plot()}
function. The example in Figure~5 plots the forecasts for Poland with a
68\% forecast interval, with optional plot's colour, main title, and
axis label.

\begin{CodeChunk}
\begin{CodeInput}
R> fore |>    
+   plot( 
+     which_c = "POL",  
+     probability = 0.68, 
+     col = "#1A003F",
+     main = "Labour Market Forecasts for Poland", 
+     xlab = "time [years]"
+   )
\end{CodeInput}
\begin{figure}

{\centering \includegraphics{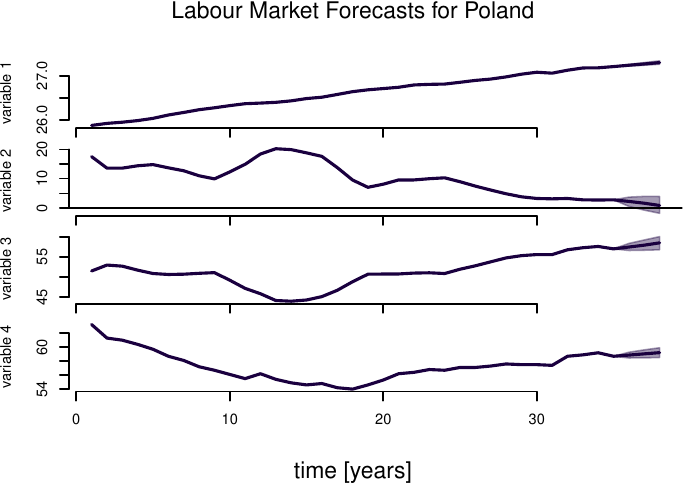} 

}

\caption[Three years ahead forecasts for Poland]{Three years ahead forecasts for Poland}\label{fig:fore_plo_both}
\end{figure}
\end{CodeChunk}

The documentation for the \texttt{summary()} and \texttt{plot()}
functions can be accessed by running \texttt{?summary.ForecastsPANEL}
and \texttt{?plot.ForecastsPANEL}.

\subsection{Work with a Model with Constrained Forecasts}

The results reported in Figure~5 include unconstrained forecasts. The
interval forecasts of unemployment rate, \(UR\), include negative
values, which makes the forecasts difficult to interpret and report in
official communication. The labour market rates are variables specified
within the interval from 0 to 100. The package facilitates constrained
forecasting of rates ensuring the draws from the predictive density fall
within the interval. To apply the constraint, use the argument
\texttt{type} in the \texttt{specify\_bvarPANEL} function by setting it
to \texttt{c("real",\ "rate",\ "rate",\ "rate")}. This setup results in
unconstrained forecasts for the first variable, \(gdp\), and treats the
remaining three, that is \(UR\), \(EPR\), and \(LFPR\), as rates. Follow
by estimating the model.

\begin{CodeChunk}
\begin{CodeInput}
R> ilo_dynamic_panel |> 
+   specify_bvarPANEL$new(
+     exogenous = ilo_exogenous_variables,
+     type = c("real", "rate", "rate", "rate")
+   ) |> 
+   estimate(S = 1000, show_progress = FALSE) |> 
+   estimate(S = 1000, show_progress = FALSE) -> post_cf
\end{CodeInput}
\end{CodeChunk}

In the second step, predict and plot the forecasts in Figure~6.
Unemployment rate forecasts are now bind by the restrictions. The
variable type can be specified for all the models in the package.

\begin{CodeChunk}
\begin{CodeInput}
R> post_cf |>  
+   forecast( 
+     horizon = 3,
+     exogenous_forecast = ilo_exogenous_forecasts 
+   ) |> 
+   plot( 
+     which_c = "POL",  
+     probability = 0.68, 
+     main = "Labour Market Forecasts for Poland", 
+     xlab = "time [years]"
+   )
\end{CodeInput}
\begin{figure}

{\centering \includegraphics{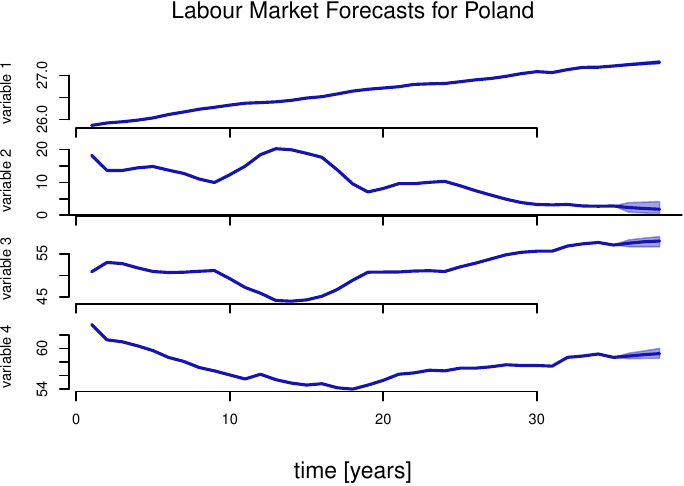} 

}

\caption[Three years ahead forecasts for Poland]{Three years ahead forecasts for Poland: forecasted values for rates fall within interval from 0 to 100}\label{fig:fore1cf}
\end{figure}
\end{CodeChunk}

\subsection{Forecast Conditionally, Given Future Values of Other Variables}

The package facilitates conditional forecasting given the future values
of some variables. This feature is illustrated by forecasting labour
market outcomes three years ahead given the future values of gross
domestic product. The package includes an object
\texttt{ilo\_conditional\_forecasts} with these future values
appropriately formatted as a list of matrices. Each of these matrices
includes the number of rows equal to the forecast horizon and the number
of columns equal to the number of dependent variables. Its missing
values coded using \texttt{NA} denote the future values to be
forecasted, while any provided numerical values are conditioned on in
the forecasting. The listing below is an example of future values of
Polish gross domestic product used to conditionally forecast labour
market outcomes.

\begin{CodeChunk}
\begin{CodeInput}
R> ilo_conditional_forecasts$POL
\end{CodeInput}
\begin{CodeOutput}
Time Series:
Start = 2025 
End = 2027 
Frequency = 1 
          gdp UR EPR LFPR
2025 27.24389 NA  NA   NA
2026 27.27435 NA  NA   NA
2027 27.30252 NA  NA   NA
\end{CodeOutput}
\end{CodeChunk}

To use the future values in forecasting, provide the list object
\texttt{ilo\_conditional\_forecasts} as the value for argument
\texttt{conditional\_forecast} of the \texttt{forecast} method.

\begin{CodeChunk}
\begin{CodeInput}
R> post_cf |>  
+   forecast( 
+     horizon = 3,
+     exogenous_forecast = ilo_exogenous_forecasts ,
+     conditional_forecast = ilo_conditional_forecasts
+   ) |> 
+   plot( 
+     which_c = "POL",  
+     probability = 0.68, 
+     main = "Labour Market Forecasts for Poland", 
+     xlab = "time [years]"
+   )
\end{CodeInput}
\begin{figure}

{\centering \includegraphics{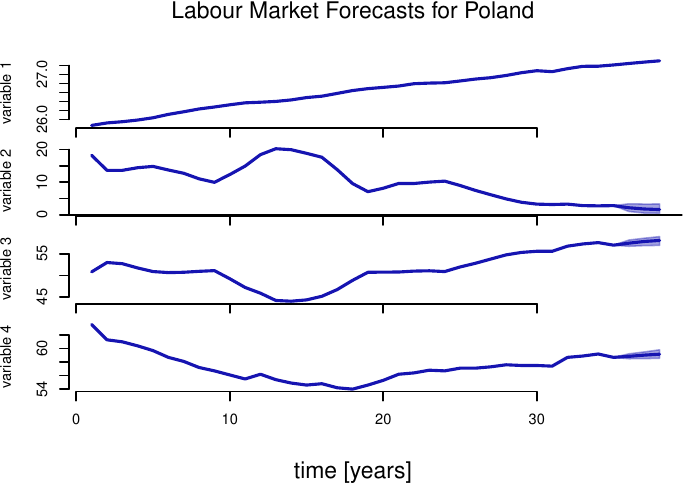} 

}

\caption[Three years ahead forecasts for Poland]{Three years ahead forecasts for Poland: labour market outcomes are forecasted given the future values for gdp}\label{fig:fore1codf}
\end{figure}
\end{CodeChunk}

The plot of the conditional forecast also includes the provided future
values.

\section{Perform Pseudo-out-of-sample Forecasting Exercise}\label{sec:poosf}

The \pkg{bpvars} package provides a range of functions to perform a
pseudo-out-of-sample forecasting exercise and report forecasting
performance measures. Begin, by specifying a model using the function
\texttt{specify\_bvarPANEL\$new()} or
\texttt{specify\_bvarGroupPANEL\$new()} as in the listing below. Then,
specify the parameters of the pseudo-out-of-sample (poos) forecasting
exercise as in the second line of the listing. In this example, we chose
the model to be estimated using \texttt{1000} draws in the burn-in run
of the MCMC algorithm to obtain convergence followed by \texttt{2000}
draws from the posterior and predictive distributions. The
\texttt{horizons} at which the forecast performance is evaluated are set
to \texttt{1}. Finally, the argument \texttt{training\_sample} set to
\texttt{20} means that the first \(T_{f_1} = 20\) annual observations
will be used to estimate the model in the first iteration, and a total
of \(F = 14\) iterations will be performed and used to evaluate the
forecasts and predictive ability of models. Both the specification of
the model object \texttt{spec} and the specification of the forecasting
exercise \texttt{poos} are provided as arguments to function
\texttt{forecast\_poos\_recursively()}, which runs the
pseudo-out-of-sample forecasting exercise and returns the forecasts in
object \texttt{fore}. Depending on the setup of the exercise, the
execution of this function might take substantial computational
resources and time as it estimates the model thirteen times. One
full-sample estimation is used to provide adequate starting values for
the estimations in the following fourteen iterations of the exercise.

\begin{CodeChunk}
\begin{CodeInput}
R> spec = specify_bvarPANEL$new(ilo_dynamic_panel)
R> poos = specify_poosf_exercise$new(
+           spec, 
+           S = 2000, 
+           S_burn = 1000, 
+           horizons = 1, 
+           training_sample = 20
+         )
R> fore = forecast_poos_recursively(spec, poos)
\end{CodeInput}
\begin{CodeOutput}
**************************************************|
 bpvars: Forecasting with Bayesian Panel VARs     |
**************************************************|
 Recursive pseudo-out-of-sample forecasting using
         expanding window samples.
 Press Esc to interrupt the computations
**************************************************|
 Step 1: Estimate a model for a full sample to get
         starting values for subsequent steps.
 Step 2: Recursive pseudo out-of-sample
         forecasting performed for 14 samples.
**************************************************|
\end{CodeOutput}
\end{CodeChunk}

Finally, one line computes the forecast performance measures using the
function \texttt{compute\_forecast\_performance()} and saves them in
object \texttt{fper}.

\begin{CodeChunk}
\begin{CodeInput}
R> fper = compute_forecast_performance(fore)
\end{CodeInput}
\end{CodeChunk}

Two points are due at this stage. Firstly, the forecasts from the
pseudo-out-of sample exercise are of class \texttt{"ForecastsPANEL"} and
can be as such analysed and visualised using the \texttt{summary()} and
\texttt{plot()} methods. For instance, the object \texttt{fore} includes
forecasts for Poland from the fourth iteration of the forecasting
exercise:

\begin{CodeChunk}
\begin{CodeInput}
R> class(fore[[4]]$POL)
\end{CodeInput}
\begin{CodeOutput}
[1] "Forecasts"
\end{CodeOutput}
\end{CodeChunk}

Secondly, one can analyse the forecasting performance of this particular
model by looking at the computed performance measures. For instance, the
line below displays the overall predictive log-scores for the Panel VAR
model

\begin{CodeChunk}
\begin{CodeInput}
R> fper$PLS$Global
\end{CodeInput}
\begin{CodeOutput}
               1
gdp    1.9410814
UR    -0.1257684
EPR   -0.4785088
LFPR  -0.1318001
joint  4.2588116
\end{CodeOutput}
\end{CodeChunk}

whereas the following line displays, root-mean-squared-forecast error
for Poland.

\begin{CodeChunk}
\begin{CodeInput}
R> fper$RMSFE$POL
\end{CodeInput}
\begin{CodeOutput}
               1
gdp   0.03257984
UR    0.48973107
EPR   0.52115363
LFPR  0.37134173
joint 0.40323497
\end{CodeOutput}
\end{CodeChunk}

We now move on to practical implementation of the forecast performance
comparison with other models, namely, Panel VARs with country grouping
where the group allocations are driven by geographical location,

\begin{CodeChunk}
\begin{CodeInput}
R> spec_g = specify_bvarGroupPANEL$new(
+             ilo_dynamic_panel,
+             group_allocation = country_grouping_region
+          )
R> poos_g = specify_poosf_exercise$new(
+             spec_g, 
+             S = 2000, 
+             S_burn = 1000, 
+             horizons = 1, 
+             training_sample = 20
+          )
R> fore_g = forecast_poos_recursively(spec_g, poos_g, show_progress = FALSE)
R> fper_g = compute_forecast_performance(fore_g)
\end{CodeInput}
\end{CodeChunk}

or by the income group.

\begin{CodeChunk}
\begin{CodeInput}
R> spec_ge = specify_bvarGroupPANEL$new(
+             ilo_dynamic_panel,
+             group_allocation = country_grouping_incomegroup
+          )
R> poos_ge = specify_poosf_exercise$new(
+             spec_ge, 
+             S = 2000, 
+             S_burn = 1000, 
+             horizons = 1, 
+             training_sample = 20
+          )
R> fore_ge = forecast_poos_recursively(spec_ge, poos_ge, show_progress = FALSE)
R> fper_ge = compute_forecast_performance(fore_ge)
\end{CodeInput}
\end{CodeChunk}

Below, we report the relative predictive log-scores one period ahead for
the models with country groupings, in the order of mentioning them above
in columns, to the model with country-specific parameters. The negative
signs of these values indicate a higher \(PLS\) value for the latter
model.

\begin{CodeChunk}
\begin{CodeInput}
R> cbind(
+   fper_g$PLS$Global[,1] - fper$PLS$Global[,1],
+   fper_ge$PLS$Global[,1] - fper$PLS$Global[,1]
+ )
\end{CodeInput}
\begin{CodeOutput}
            [,1]       [,2]
gdp   -0.4642570 -0.6179373
UR    -0.9915019 -1.0670780
EPR   -1.5368259 -1.8567238
LFPR  -1.8820109 -2.1735634
joint -4.0593358 -4.5932529
\end{CodeOutput}
\end{CodeChunk}

In another example we check which of the models systematically predicts
better the outcomes for Poland one year ahead. Therefore, we report the
relative root-mean-squared-forecast error of the models with country
groupings to the model with country-specific parameters.

\begin{CodeChunk}
\begin{CodeInput}
R> cbind(
+   fper_g$RMSFE$POL[,1]/fper$RMSFE$POL[,1],
+   fper_ge$RMSFE$POL[,1]/fper$RMSFE$POL[,1]
+ )
\end{CodeInput}
\begin{CodeOutput}
           [,1]     [,2]
gdp   0.6578034 1.416058
UR    0.5826982 0.740078
EPR   1.6657652 3.106345
LFPR  3.0037354 4.956151
joint 1.7881705 3.072889
\end{CodeOutput}
\end{CodeChunk}

These values being greater than one for all outcomes show that the
country-specific Panel VAR also predicts better in terms of point
forecast for Poland.

\section{Work with Missing Observations}\label{sec:missing}

Working with macroeconomic dynamic panel data often requires handling
missing observations. The data set from object
\texttt{ilo\_dynamic\_panel} used in the previous sections includes
cubic data in which the missing observations are replaced by the values
imputed by the Statistics Department at the International Labour
organisation. For the sake of illustrating the handling of missing
observations in the \pkg{bpvars} package, we estimate the model using
data in object \texttt{ilo\_dynamic\_panel\_missing} containing just
over 34\% of missing observations. It is formatted the same way as
object \texttt{ilo\_dynamic\_panel}, but it contains \texttt{NA} for
missing observations. The series are plotted in Figure~8.

\begin{CodeChunk}
\begin{figure}

{\centering \includegraphics{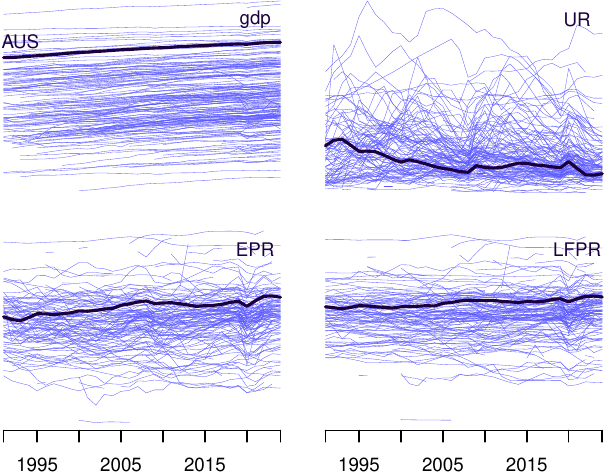} 

}

\caption[Dynamic panel data with missing observations]{Dynamic panel data with missing observations}\label{fig:missingdataplot}
\end{figure}
\end{CodeChunk}

In what follows, we proceed with the usual specification and estimation
of the model. The data object \texttt{ilo\_dynamic\_panel\_missing} is
provided as the first argument to the function specifying the model
\texttt{specify\_bvarPANEL}. This function extracts the information
regarding missing data and provides it in an appropriate form to the
\texttt{estimate} function. Any model in the package can handle data
objects with missing observations.

\begin{CodeChunk}
\begin{CodeInput}
R> ilo_dynamic_panel_missing |> 
+   specify_bvarPANEL$new() |> 
+   estimate(S = 1000, show_progress = FALSE) |> 
+   estimate(S = 1000, show_progress = FALSE) -> post_miss
\end{CodeInput}
\end{CodeChunk}

For illustrative purposes, Figure~9 reports Brazilian unemployment rate
that includes missing observations for 1994, 2000, and 2010. In
Figure~9, the black thin line represents the posterior mean of the
estimated missing observations and the blue thin line those from object
\texttt{ilo\_dynamic\_panel}. Despite some differences, these
observations follow quite closely.

\begin{CodeChunk}
\begin{CodeInput}
R> ur_bra = ts(
+   apply(post_miss$posterior$Y[22,1][[1]],1:2,mean)[,2], 
+   start = c(1991), frequency = 1
+ )
R> plot(
+   ilo_dynamic_panel$BRA[,2], 
+   col = "#5A58FF",  bty = "n", ylab = "", 
+   main = "Unemployment rate in Brasil"
+ )
R> lines(ur_bra)
R> lines(ilo_dynamic_panel_missing$BRA[,2], col = "#1A003F", lwd = 2)
R> legend(1991, 14, 
+        legend = c("estimated", "imputed", "original data"), 
+        col = c("black", "#5A58FF", "#1A003F"), lwd = c(1, 1, 2), bty = "n"
+ ) 
\end{CodeInput}
\begin{figure}

{\centering \includegraphics{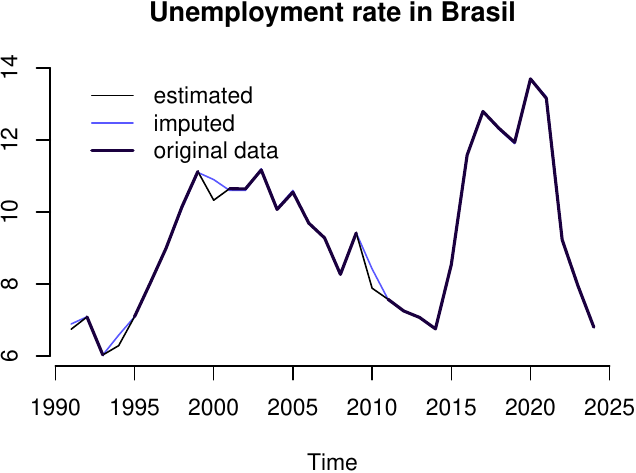} 

}

\caption[Illustration of estimated and imputed missing observations for the unemployment rate in Brazil]{Illustration of estimated and imputed missing observations for the unemployment rate in Brazil}\label{fig:plot1miss}
\end{figure}
\end{CodeChunk}

Finally, we illustrate the effect of missing observations on forecasts
on the example of Poland. For this country, the data begins in 1992.
However, its forecasts are affected by the uncertainty due to missing
observations also in other countries. This is reflected in slightly
wider predictive intervals.

\begin{CodeChunk}
\begin{CodeInput}
R> post_miss |>
+   forecast(horizon = 3) |>
+   plot(
+     which_c = "POL",
+     probability = 0.68,
+     main = "Labour Market Forecasts for Poland",
+     xlab = "time [years]"
+   )
\end{CodeInput}
\begin{figure}

{\centering \includegraphics{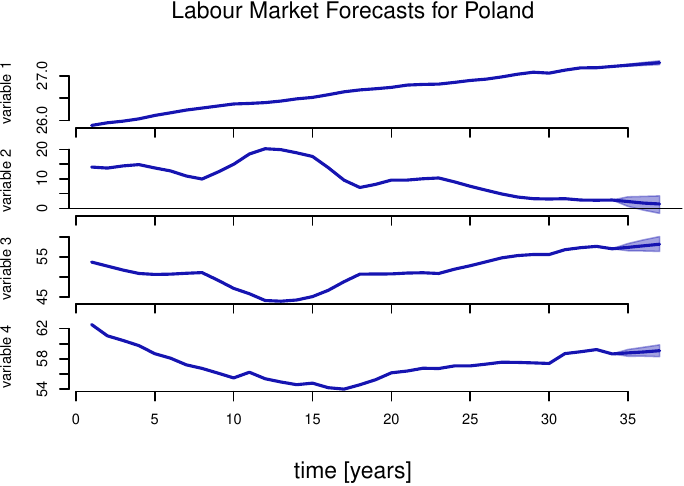} 

}

\caption[Three years ahead forecasts for Poland]{Three years ahead forecasts for Poland}\label{fig:fore1miss}
\end{figure}
\end{CodeChunk}

\section{Conclusion}

We propose a new set of forecasting models for labour market outcomes
that are based on newest modelling practices and inspired by solutions
implemented for global forecasting for United Nation Agencies. This
framework produces reliable and interpretable forecasts, as well as a
range of other tools supporting diversification and accountability of
risks in the process of advising and decision-making. These tools
include convenient forecasting reporting and visualisation, analysis of
forecast error variance decompositions, and pseudo-out-of-sample
forecasting exercises for the forecasting performance assessment.

The models are implemented in the \proglang{R} package \pkg{bpvars} that
is available on CRAN. The package provides a simple workflow for the
model specification, estimation, forecasting, and their summaries and
visualisations. The package is designed to be user-friendly and
accessible to a wide range of users, including those with limited
experience in Bayesian econometrics. The package builds on the best of
the two worlds: excellent computational speed thanks to algorithms
written in \proglang{C++} and the convenience of data analysis in
\proglang{R}. Both of these features greatly simplify the work with the
Bayesian dynamic hierarchical system models excelling in forecasting
performance.

\appendix
\section{Bayesian Estimation of the Model}\label{sec:gibbs}

\noindent The estimation of the model proceeds by Gibbs sampler
\citep[see][]{casella_explaining_1992} that is a numerical procedure of
obtaining a sample of random draws from the joint posterior distribution
of the parameters of the model and missing observations given data by
sampling from the full conditional posterior distributions for
individual groups of parameters. Below, the details of the particular
full conditional posterior distributions for parameters are provided.
This is complemented by the full conditional posterior distribution of
missing observations presented in Section \ref{sec:miss}.

\subsection{Sampling Country-Specific Parameters}

\noindent The natural-conjugate prior distributions for country-specific
parameters results in the joint full conditional posterior distribution
for the autoregressive parameters and the error term variances have the
convenient form of the matrix-variate normal inverse-Wishart
distribution. \begin{align}
\mathbf{A}_c, \boldsymbol\Sigma_c | \mathbf{Y}_c, \mathbf{X}_c, \mathbf{A}, \mathbf{V}, \mathbf{\Sigma}, \nu &\sim\mathcal{MNIW}_{K\times N}\left(\overline{\mathbf{A}}, \overline{\mathbf{V}}, \overline{\mathbf{\Sigma}}, \overline{\nu}\right)\\
\overline{\mathbf{V}} &= \left[ \mathbf{X}_c'\mathbf{X}_c + \mathbf{V}^{-1}\right]^{-1}\\
\overline{\mathbf{A}} &= \overline{\mathbf{V}} \left[ \mathbf{X}_c'\mathbf{Y}_c + \mathbf{V}^{-1}\mathbf{A}\right]\\
\overline{\mathbf{\Sigma}} &= (\nu-N-1)\mathbf{\Sigma} + \mathbf{Y}_c'\mathbf{Y}_c + \mathbf{A}'\mathbf{V}^{-1}\mathbf{A} - \overline{\mathbf{A}}'\overline{\mathbf{V}}^{-1}\overline{\mathbf{A}}\\
\overline{\nu} &= T_c + \nu
\end{align}

\subsection{Sampling Global Parameters}

\noindent It occurs that with the current model specification the joint
full conditional posterior distribution for the global autoregressive
slopes and their column-specific covariance is the matrix-variate normal
inverse-Wishart distribution. This is a new result not known in the
literature.

For the simplicity of notation define collections of country-specific
objects. Let
\(\mathbf{Y}_{1-C} = \left\{\mathbf{Y}_1, \dots, \mathbf{Y}_C\right\}\),
\(\mathbf{X}_{1-C} = \left\{\mathbf{X}_1, \dots, \mathbf{X}_C\right\}\),
\(\mathbf{A}_{1-C} = \left\{\mathbf{A}_1, \dots, \mathbf{A}_C\right\}\),
and
\(\mathbf{\Sigma}_{1-C} = \left\{\mathbf{\Sigma}_1, \dots, \mathbf{\Sigma}_C\right\}\).
Then the full conditional posterior distribution is specified as
\begin{align}
\mathbf{A}', \mathbf{V} | \mathbf{Y}_{1-C}, \mathbf{X}_{1-C}, \mathbf{A}_{1-C}, \mathbf{\Sigma}_{1-C} &\sim\mathcal{MNIW}_{N\times K}\left(\overline{\mathbf{M}}', \overline{\mathbf{W}}, \overline{\mathbf{S}}, \overline{\eta}\right)\\
\overline{\mathbf{S}} &= \left[\frac{1}{s}\underline{\mathbf{S}}^{-1} + \sum_{c=1}^{C}\mathbf{\Sigma}_c^{-1}\right]^{-1}\\
\overline{\mathbf{M}}' &= \overline{\mathbf{S}}\left[\frac{m}{s}\underline{\mathbf{S}}^{-1}\underline{\mathbf{M}}' + \sum_{c=1}^{C}\mathbf{\Sigma}_c^{-1}\mathbf{A}_c'\right]\\
\overline{\mathbf{W}} &= w\underline{\mathbf{W}} + \frac{m^2}{s}\underline{\mathbf{M}}\underline{\mathbf{S}}^{-1}\underline{\mathbf{M}}' + \left[\sum_{c=1}^{C} \mathbf{A}_c \mathbf{\Sigma}_c^{-1}\mathbf{A}_c'\right] - \overline{\mathbf{M}}\overline{\mathbf{S}}^{-1}\overline{\mathbf{M}}'\\
\overline{\eta} &= CN + \underline{\eta}
\end{align}

Furthermore, the global error term covariance matrix is sampled from
Wishart full conditional posterior distribution. \begin{align}
\mathbf{\Sigma} | \mathbf{Y}_{1-C}, \mathbf{X}_{1-C}, \mathbf{A}_{1-C}, \mathbf{\Sigma}_{1-C} &\sim\mathcal{W}_{N}\left(\overline{\mathbf{S}}_\Sigma, \overline{\mu}\right)\\
\overline{\mathbf{S}}_\Sigma &= \left[\frac{1}{s}\underline{\mathbf{S}}_\Sigma^{-1} + (\nu -N-1)\sum_{c=1}^{C}\mathbf{\Sigma}_c^{-1}\right]^{-1}\\
\overline{\mu} &= C\nu + \underline{\mu}_\Sigma
\end{align}

The posterior kernel of the shape parameter given in expression
\eqref{eq:nuk} does not resemble any of known distributions.
\begin{multline}
p\left(\nu|\mathbf{Y}_{1-C}, \mathbf{X}_{1-C}, \mathbf{A}_{1-C}, \mathbf{\Sigma}_{1-C}, \dots \right) \propto 2^{-\frac{CN(K+\nu)}{2}} (\nu - N - 1)^{\frac{CN\nu}{2}}\det\left(\mathbf\Sigma\right)^{\frac{C\nu}{2}}\exp\left(-\underline{\lambda}\nu\right)\\
\times
\exp\left\{-\frac{\nu-N-1}{2}\mathop{\mathrm{tr}}\left[\boldsymbol\Sigma\left(\sum_{c=1}^{C}\boldsymbol\Sigma_c^{-1}\right)\right]\right\}
\left[\prod_{n=1}^N \Gamma\left(\frac{\nu+1-n}{2}\right)\right]^{-C}
\left[\prod_{c=1}^C \det\left(\mathbf\Sigma_c\right)\right]^{-\frac{\nu + N+K+1}{2}}\label{eq:nuk}
\end{multline} Therefore, at each \(s\)th iteration, it is sampled using
Adaptive Metropolis-Hastings \citep[see][]{vihola2012} within Gibbs
strategy with the candidate generating density set to a truncated normal
distribution: \begin{align}
\nu^* &\sim\mathcal{N}\left(\nu^{(s-1)}, \sigma_s^2 Cov[\nu^{(s-1)}]\right)
\end{align} with the mean set to the current state of the Markov chain
and the variance set to the product of the adaptive scaling constant
following the dynamic rune \begin{align}
\log\sigma_s = \log\sigma_{s-1} + \frac{1}{2}\log\left(1 + s^{-0.6}(\alpha_{s} - 0.4)\right)
\end{align} where \(\alpha_{s}\) is the Metropolis acceptance
probability, and the negative inverse of the Hessian of the posterior
kernel given by \begin{align}
Cov[\nu] &= \frac{4}{CN}\left[\frac{4(N+1) - 2\nu}{(N+1-\nu)^2} + \frac{1}{N}\sum_{n=1}^{N}\psi^{(1)}\left(\frac{\nu+1-n}{2}\right)\right]^{-1}
\end{align} where \(\psi^{(1)}()\) is a poly-gamma function. The
truncation on this density reflects the constraint from specification in
\eqref{eq:csmnivprior}, namely, \(\nu>N+1\). It is assumed not to affect
the symmetry of the full conditional posterior distribution for \(\nu\).

\subsection{Sampling level-3 priors}

\noindent The average global persistence hyper-parameter is sampled from
a normal full conditional posterior distribution. \begin{align}
m \mid \mathbf{Y}_{1-C}, \mathbf{X}_{1-C}, \mathbf{A}_{1-C}, \mathbf{\Sigma}_{1-C}, \dots &\sim\mathcal{N}\left(\overline{\mu}_m,\overline{\sigma}_m^2\right)\\
\overline{\sigma}_m^2 &= \left[\underline{\sigma}_m^{-2} + 
\text{vec}(\underline{\mathbf{M}})'\left(\frac{1}{s}\underline{\mathbf{S}}^{-1} \otimes \frac{1}{w}\underline{\mathbf{V}}^{-1} \right)\text{vec}(\underline{\mathbf{M}})
\right]^{-1}
\\
\overline{\mu}_m &= \overline{\sigma}_m^2 \left[\underline{\sigma}_m^{-2}\underline{\mu}_m + 
\text{vec}(\underline{\mathbf{M}})'\left(\frac{1}{s}\underline{\mathbf{S}}^{-1} \otimes \frac{1}{w}\underline{\mathbf{V}}^{-1} \right)\text{vec}(\mathbf{A})
\right]
\end{align} The global autoregressive equation-specific level of
shrinkage is sampled from the gamma full conditional posterior
distribution. \begin{align}
w \mid \mathbf{Y}_{1-C}, \mathbf{X}_{1-C}, \mathbf{A}_{1-C}, \mathbf{\Sigma}_{1-C}, \dots &\sim\mathcal{G}\left(\overline{s}_w,\overline{a}_w\right)\\
\overline{s}_w &= \underline{s}_w + \frac{1}{2}\mathop{\mathrm{tr}}\left[\mathbf{V}^{-1}\underline{\mathbf{W}}\right]\\
\overline{a}_w &= \underline{a}_w + \frac{1}{2}\underline{\eta}K
\end{align} Finally, the global autoregressive row-specific level of
shrinkage is sampled from the inverted gamma 2 full conditional
posterior distribution. \begin{align}
s \mid \mathbf{Y}_{1-C}, \mathbf{X}_{1-C}, \mathbf{A}_{1-C}, \mathbf{\Sigma}_{1-C}, \dots &\sim\mathcal{IG}2\left(\overline{s}_s,\overline{\nu}_s\right)\\
\overline{s}_s &= \underline{s}_s + \mathop{\mathrm{tr}}\left[\mathbf{V}^{-1}\left(\mathbf{A}' - m\underline{\mathbf{M}}'\right)'\underline{\mathbf{S}}^{-1}\left(\mathbf{A}' - m\underline{\mathbf{M}}'\right)\right] + \mathop{\mathrm{tr}}\left[\underline{\mathbf{S}}_\Sigma^{-1}\mathbf{\Sigma}\right]\\
\overline{\nu}_s &= \underline{\nu}_s + KN + N\underline{\mu}_\Sigma
\end{align}

\bibliography{bib}

\begin{thebibliography}{58}
\newcommand{\enquote}[1]{``#1''}
\providecommand{\natexlab}[1]{#1}
\providecommand{\url}[1]{\texttt{#1}}
\providecommand{\urlprefix}{URL }
\expandafter\ifx\csname urlstyle\endcsname\relax
  \providecommand{\doi}[1]{doi:\discretionary{}{}{}#1}\else
  \providecommand{\doi}{doi:\discretionary{}{}{}\begingroup
  \urlstyle{rm}\Url}\fi
\providecommand{\eprint}[2][]{\url{#2}}

\bibitem[{Bauwens \emph{et~al.}(1999)Bauwens, Lubrano, and Richard}]{blr1999}
Bauwens L, Lubrano M, Richard JF (1999).
\newblock \emph{Bayesian {{Inference}} in {{Dynamic Econometric Models}}}.
\newblock Advanced {{Texts}} in {{Econometrics}}. Oxford University Press,
  Oxford.
\newblock \doi{10.1093/acprof:oso/9780198773122.001.0001}.

\bibitem[{Boeck \emph{et~al.}(2022)Boeck, Feldkircher, and
  Huber}]{boeck_bgvar_2022}
Boeck M, Feldkircher M, Huber F (2022).
\newblock \enquote{\textbf{{BGVAR}} : Bayesian Global Vector Autoregressions
  with Shrinkage Priors in \textit{R}.}
\newblock \emph{Journal of Statistical Software}, \textbf{104}(9).
\newblock \doi{10.18637/jss.v104.i09}.

\bibitem[{Boeck \emph{et~al.}(2025)Boeck, Feldkircher, and Huber}]{BGVAR}
Boeck M, Feldkircher M, Huber F (2025).
\newblock \emph{{BGVAR}: {B}ayesian Global Vector Autoregressions}.
\newblock {R} package version 2.5.9,
  \urlprefix\url{https://CRAN.R-project.org/package=BGVAR}.

\bibitem[{Botev and Belzile(2024)}]{TruncatedNormal}
Botev Z, Belzile L (2024).
\newblock \emph{TruncatedNormal: Truncated Multivariate Normal and Student
  Distributions}.
\newblock \doi{10.32614/CRAN.package.TruncatedNormal}.
\newblock R package version 2.3,
  \urlprefix\url{https://CRAN.R-project.org/package=TruncatedNormal}.

\bibitem[{Botev(2016)}]{Botev}
Botev ZI (2016).
\newblock \enquote{The Normal Law Under Linear Restrictions: Simulation and
  Estimation via Minimax Tilting.}
\newblock \emph{Journal of the Royal Statistical Society Series B: Statistical
  Methodology}, \textbf{79}(1), 125--148.
\newblock \doi{10.1111/rssb.12162}.

\bibitem[{Canova(2007)}]{Canova2007}
Canova F (2007).
\newblock \emph{Methods for Applied Macroeconomic Research}.
\newblock Princeton University Press, Princeton.
\newblock ISBN 9781400841028.
\newblock \doi{doi:10.1515/9781400841028}.
\newblock \urlprefix\url{https://doi.org/10.1515/9781400841028}.

\bibitem[{Canova and Ciccarelli(2013)}]{canova2013}
Canova F, Ciccarelli M (2013).
\newblock \enquote{Panel Vector Autoregressive Models: A Survey.}
\newblock In TB~Fomby, L~Kilian, A~Murphy (eds.), \emph{{VAR} Models in
  Macroeconomics – New Developments and Applications: Essays in Honor of
  Christopher A. Sims}, volume~32, pp. 205--246. Emerald Group Publishing
  Limited.
\newblock \doi{10.1108/S0731-9053(2013)0000031006}.
\newblock Series Title: Advances in Econometrics.

\bibitem[{Casella and George(1992)}]{casella_explaining_1992}
Casella G, George EI (1992).
\newblock \enquote{Explaining the Gibbs Sampler.}
\newblock \emph{The American Statistician}, \textbf{46}(3), 167--174.
\newblock \doi{10.1080/00031305.1992.10475878}.

\bibitem[{Chang(2021)}]{R6}
Chang W (2021).
\newblock \emph{\pkg{R6}: Encapsulated Classes with Reference Semantics}.
\newblock R package version 2.5.1,
  \urlprefix\url{https://CRAN.R-project.org/package=R6}.

\bibitem[{Croissant and Millo(2008)}]{plm_pap}
Croissant Y, Millo G (2008).
\newblock \enquote{Panel Data Econometrics in {R}: The {plm} Package.}
\newblock \emph{Journal of Statistical Software}, \textbf{27}(2), 1--43.
\newblock \doi{10.18637/jss.v027.i02}.

\bibitem[{Croissant \emph{et~al.}(2025)Croissant, Millo, and Tappe}]{plm}
Croissant Y, Millo G, Tappe K (2025).
\newblock \emph{plm: Linear Models for Panel Data}.
\newblock \doi{10.32614/CRAN.package.plm}.
\newblock R package version 2.6-7,
  \urlprefix\url{https://CRAN.R-project.org/package=plm}.

\bibitem[{Cuaresma \emph{et~al.}(2016)Cuaresma, Feldkircher, and
  Huber}]{CFH2016}
Cuaresma JC, Feldkircher M, Huber F (2016).
\newblock \enquote{Forecasting with Global Vector Autoregressive Models: a
  Bayesian Approach.}
\newblock \emph{Journal of Applied Econometrics}, \textbf{31}(7), 1371--1391.
\newblock \doi{https://doi.org/10.1002/jae.2504}.

\bibitem[{Dagum and Menon(1998)}]{dagum1998openmp}
Dagum L, Menon R (1998).
\newblock \enquote{OpenMP: an industry standard API for shared-memory
  programming.}
\newblock \emph{Computational Science \& Engineering, IEEE}, \textbf{5}(1),
  46--55.

\bibitem[{{David Bescond}(2024)}]{Rilostat}
{David Bescond} (2024).
\newblock \emph{Rilostat: ILO Open Data via Ilostat Bulk Download Facility}.
\newblock \doi{10.32614/CRAN.package.Rilostat}.
\newblock R package version 2.1.0,
  \urlprefix\url{https://CRAN.R-project.org/package=Rilostat}.

\bibitem[{Dieppe \emph{et~al.}(2016)Dieppe, Legrand, and van Roye}]{Dieppe2016}
Dieppe A, Legrand R, van Roye B (2016).
\newblock \enquote{The BEAR toolbox.}
\newblock \emph{ECB Working Paper}, \textbf{1934}, 1--291.
\newblock \doi{https://doi.org/10.2866/292952}.

\bibitem[{Dieppe and van Roye(2024)}]{BEAR}
Dieppe A, van Roye B (2024).
\newblock \emph{The BEAR toolbox}.
\newblock MatLab library version 5.2.2,
  \urlprefix\url{https://github.com/european-central-bank/BEAR-toolbox/}.

\bibitem[{Director \emph{et~al.}(2021)Director, Raftery, and
  Bitz}]{director_probabilistic_2021}
Director HM, Raftery AE, Bitz CM (2021).
\newblock \enquote{Probabilistic forecasting of the Arctic sea ice edge with
  contour modeling.}
\newblock \emph{The Annals of Applied Statistics}, \textbf{15}(2).
\newblock \doi{10.1214/20-AOAS1405}.

\bibitem[{Doan \emph{et~al.}(1984)Doan, Litterman, and Sims}]{Doan1984}
Doan T, Litterman RB, Sims CA (1984).
\newblock \enquote{{Forecasting and Conditional Projection Using Realistic
  Prior Distributions}.}
\newblock \emph{Econometric Reviews}, \textbf{3}(1), 1--100.
\newblock \doi{10.1080/07474938408800053}.

\bibitem[{Eddelbuettel(2013)}]{eddelbuettel_seamless_2013}
Eddelbuettel D (2013).
\newblock \emph{Seamless {R} and {C}++ {Integration} with \pkg{Rcpp}}.
\newblock Springer, New York, NY.
\newblock \doi{10.1007/978-1-4614-6868-4}.

\bibitem[{Eddelbuettel \emph{et~al.}(2011)Eddelbuettel, Fran{\c{c}}ois,
  Allaire, Ushey, Kou, Russel, Chambers, and Bates}]{eddelbuettel2011rcpp}
Eddelbuettel D, Fran{\c{c}}ois R, Allaire J, Ushey K, Kou Q, Russel N, Chambers
  J, Bates D (2011).
\newblock \enquote{\pkg{Rcpp}: Seamless {R} and {C++} integration.}
\newblock \emph{Journal of Statistical Software}, \textbf{40}(8), 1--18.
\newblock \doi{10.18637/jss.v040.i08}.

\bibitem[{Eddelbuettel \emph{et~al.}(2025{\natexlab{a}})Eddelbuettel, Francois,
  Allaire, Ushey, Kou, Russell, Ucar, Bates, and Chambers}]{Rcpp}
Eddelbuettel D, Francois R, Allaire J, Ushey K, Kou Q, Russell N, Ucar I, Bates
  D, Chambers J (2025{\natexlab{a}}).
\newblock \emph{Rcpp: Seamless R and C++ Integration}.
\newblock \doi{10.32614/CRAN.package.Rcpp}.
\newblock R package version 1.1.0,
  \urlprefix\url{https://CRAN.R-project.org/package=Rcpp}.

\bibitem[{Eddelbuettel \emph{et~al.}(2025{\natexlab{b}})Eddelbuettel, Francois,
  Bates, Ni, and Sanderson}]{RcppArmadillo}
Eddelbuettel D, Francois R, Bates D, Ni B, Sanderson C (2025{\natexlab{b}}).
\newblock \emph{RcppArmadillo: 'Rcpp' Integration for the 'Armadillo' Templated
  Linear Algebra Library}.
\newblock \doi{10.32614/CRAN.package.RcppArmadillo}.
\newblock R package version 15.0.2-2,
  \urlprefix\url{https://CRAN.R-project.org/package=RcppArmadillo}.

\bibitem[{Eddelbuettel and Sanderson(2014)}]{eddelbuettel_rcpparmadillo2014}
Eddelbuettel D, Sanderson C (2014).
\newblock \enquote{\pkg{RcppArmadillo}: {Accelerating} {R} with
  high-performance {C++} linear algebra.}
\newblock \emph{Computational Statistics \& Data Analysis}, \textbf{71},
  1054--1063.
\newblock \doi{10.1016/j.csda.2013.02.005}.

\bibitem[{Forner(2020)}]{RcppProgress}
Forner K (2020).
\newblock \emph{\pkg{RcppProgress}: An Interruptible Progress Bar with OpenMP
  Support for C++ in R Packages}.
\newblock R package version 0.4.2,
  \urlprefix\url{https://CRAN.R-project.org/package=RcppProgress}.

\bibitem[{Fosdick and Raftery(2014)}]{fosdick_regional_2014}
Fosdick B, Raftery AE (2014).
\newblock \enquote{Regional probabilistic fertility forecasting by modeling
  between-country correlations.}
\newblock \emph{Demographic Research}, \textbf{30}, 1011--1034.
\newblock \doi{10.4054/DemRes.2014.30.35}.

\bibitem[{Friedrich and Empting(2025)}]{pvars}
Friedrich L, Empting T (2025).
\newblock \emph{The {pvars} {R}-Package: VAR Modeling for Heterogeneous
  Panels}.
\newblock \doi{10.32614/CRAN.package.pvars}.
\newblock R package version 1.1.1,
  \urlprefix\url{https://CRAN.R-project.org/package=pvars}.

\bibitem[{Gelfand and Smith(1990)}]{gelfand1990sampling}
Gelfand AE, Smith AF (1990).
\newblock \enquote{Sampling-based approaches to calculating marginal
  densities.}
\newblock \emph{Journal of the American statistical association},
  \textbf{85}(410), 398--409.
\newblock \doi{10.1080/01621459.1990.10476213}.

\bibitem[{Gerland \emph{et~al.}(2014)Gerland, Raftery,
  {\v{S}}ev{\v{c}}{\'\i}kov{\'a}, Li, Gu, Spoorenberg, Alkema, Fosdick, Chunn,
  Lalic \emph{et~al.}}]{gerland2014world}
Gerland P, Raftery AE, {\v{S}}ev{\v{c}}{\'\i}kov{\'a} H, Li N, Gu D,
  Spoorenberg T, Alkema L, Fosdick BK, Chunn J, Lalic N, \emph{et~al.} (2014).
\newblock \enquote{World population stabilization unlikely this century.}
\newblock \emph{Science}, \textbf{346}(6206), 234--237.

\bibitem[{Geweke and Amisano(2010)}]{geweke_comparing_2010}
Geweke J, Amisano G (2010).
\newblock \enquote{Comparing and evaluating Bayesian predictive distributions
  of asset returns.}
\newblock \emph{International Journal of Forecasting}, \textbf{26}(2),
  216--230.
\newblock \doi{10.1016/j.ijforecast.2009.10.007}.

\bibitem[{Godwin and Raftery(2017)}]{godwin_bayesian_2017}
Godwin J, Raftery AE (2017).
\newblock \enquote{Bayesian projection of life expectancy accounting for the
  {HIV}/{AIDS} epidemic.}
\newblock \emph{Demographic Research}, \textbf{37}, 1549--1610.
\newblock \doi{10.4054/DemRes.2017.37.48}.

\bibitem[{Guttman and Menzefricke(1983)}]{guttman_bayesian_1983}
Guttman I, Menzefricke U (1983).
\newblock \enquote{Bayesian Inference in Multivariate Regression With Missing
  Observations on the Response Variables.}
\newblock \emph{Journal of Business \& Economic Statistics}, \textbf{1}(3),
  239--248.
\newblock \doi{10.1080/07350015.1983.10509347}.

\bibitem[{{International Labour Organization}(2020)}]{ILO}
{International Labour Organization} (2020).
\newblock \enquote{{ILO modelled estimates database ILOSTAT [database]}.}
\newblock \emph{Data}.
\newblock {Retrieved from ILOSTAT database [accessed May 11, 2024]},
  \urlprefix\url{https://ilostat.ilo.org/data/}.

\bibitem[{Irons and Raftery(2021)}]{irons_estimating_2021}
Irons NJ, Raftery AE (2021).
\newblock \enquote{Estimating {SARS}-{CoV}-2 infections from deaths, confirmed
  cases, tests, and random surveys.}
\newblock \emph{Proceedings of the National Academy of Sciences},
  \textbf{118}(31), e2103272118.
\newblock \doi{10.1073/pnas.2103272118}.

\bibitem[{Jarociński(2010)}]{jarocinski_responses_2010}
Jarociński M (2010).
\newblock \enquote{Responses to monetary policy shocks in the east and the west
  of Europe: a comparison.}
\newblock \emph{Journal of Applied Econometrics}, \textbf{25}(5), 833--868.
\newblock \doi{10.1002/jae.1082}.

\bibitem[{Karlsson(2013)}]{karlsson2013}
Karlsson S (2013).
\newblock \enquote{Forecasting with {{Bayesian Vector Autoregression}}.}
\newblock In \emph{Handbook of {{Economic Forecasting}}}, volume~2, pp.
  791--897. Elsevier.
\newblock \doi{10.1016/B978-0-444-62731-5.00015-4}.

\bibitem[{Lindley and Smith(1972)}]{LindleySmith1972}
Lindley DV, Smith AFM (1972).
\newblock \enquote{Bayes Estimates for the Linear Model.}
\newblock \emph{Journal of the Royal Statistical Society: Series B
  (Methodological)}, \textbf{34}(1), 1--18.
\newblock \doi{10.1111/j.2517-6161.1972.tb00885.x}.

\bibitem[{Liu and Raftery(2024)}]{liu_bayesian_2024}
Liu DH, Raftery AE (2024).
\newblock \enquote{Bayesian projections of total fertility rate conditional on
  the United Nations sustainable development goals.}
\newblock \emph{The Annals of Applied Statistics}, \textbf{18}(1).
\newblock \doi{10.1214/23-AOAS1793}.

\bibitem[{{R Core Team}(2021)}]{Rcore}
{R Core Team} (2021).
\newblock \emph{{R}: A Language and Environment for Statistical Computing}.
\newblock R Foundation for Statistical Computing, Vienna, Austria.
\newblock \urlprefix\url{https://www.R-project.org/}.

\bibitem[{Raftery \emph{et~al.}(2014{\natexlab{a}})Raftery, Alkema, and
  Gerland}]{raftery_bayesian_2014}
Raftery AE, Alkema L, Gerland P (2014{\natexlab{a}}).
\newblock \enquote{Bayesian Population Projections for the United Nations.}
\newblock \emph{Statistical Science}, \textbf{29}(1).
\newblock \doi{10.1214/13-STS419}.

\bibitem[{Raftery \emph{et~al.}(2014{\natexlab{b}})Raftery, Lalic, and
  Gerland}]{raftery_joint_2014}
Raftery AE, Lalic N, Gerland P (2014{\natexlab{b}}).
\newblock \enquote{Joint probabilistic projection of female and male life
  expectancy.}
\newblock \emph{Demographic Research}, \textbf{30}, 795--822.
\newblock \doi{10.4054/DemRes.2014.30.27}.

\bibitem[{Raftery \emph{et~al.}(2017)Raftery, Zimmer, Frierson, Startz, and
  Liu}]{raftery2017}
Raftery AE, Zimmer A, Frierson DM, Startz R, Liu P (2017).
\newblock \enquote{Less than 2 {$^\circ$}c Warming by 2100 Unlikely.}
\newblock \emph{Nature Climate Change}, \textbf{7}(9), 637--641.
\newblock \doi{10.1038/nclimate3352}.

\bibitem[{Raftery and Ševčíková(2023)}]{raftery_probabilistic_2023}
Raftery AE, Ševčíková H (2023).
\newblock \enquote{Probabilistic population forecasting: Short to very
  long-term.}
\newblock \emph{International Journal of Forecasting}, \textbf{39}(1), 73--97.
\newblock \doi{10.1016/j.ijforecast.2021.09.001}.

\bibitem[{Rendon(2013)}]{rendon2013}
Rendon SR (2013).
\newblock \enquote{Fixed and {{Random Effects}} in {{Classical}} and {{Bayesian
  Regression}}.}
\newblock \emph{Oxford Bulletin of Economics and Statistics}, \textbf{75}(3),
  460--476.
\newblock \doi{10.1111/j.1468-0084.2012.00700.x}.

\bibitem[{Sanderson and Curtin(2016)}]{sanderson2016armadillo}
Sanderson C, Curtin R (2016).
\newblock \emph{Journal of Open Source Software}, \textbf{1}(2), 26.
\newblock \doi{10.21105/joss.00026}.

\bibitem[{Sigmund and Ferstl(2021)}]{panelvar_pap}
Sigmund M, Ferstl R (2021).
\newblock \enquote{Panel Vector Autoregression in R with the package panelvar.}
\newblock \emph{The Quarterly Review of Economics and Finance}.
\newblock \doi{10.1016/j.qref.2019.01.001}.

\bibitem[{Sigmund and Ferstl(2024)}]{panelvar}
Sigmund M, Ferstl R (2024).
\newblock \emph{panelvar: Panel Vector Autoregression}.
\newblock \doi{10.32614/CRAN.package.panelvar}.
\newblock R package version 0.5.6,
  \urlprefix\url{https://CRAN.R-project.org/package=panelvar}.

\bibitem[{Sims(1980)}]{Sims1980}
Sims CA (1980).
\newblock \enquote{{Macroeconomics and Reality}.}
\newblock \emph{Econometrica}, \textbf{48}(1), 1--48.
\newblock \doi{10.2307/1912017}.

\bibitem[{Sokolov \emph{et~al.}(2009)Sokolov, Stone, Forest, Prinn, Sarofim,
  Webster, Paltsev, Schlosser, Kicklighter, Dutkiewicz, Reilly, Wang, Felzer,
  Melillo, and Jacoby}]{sokolov_probabilistic_2009}
Sokolov AP, Stone PH, Forest CE, Prinn R, Sarofim MC, Webster M, Paltsev S,
  Schlosser CA, Kicklighter D, Dutkiewicz S, Reilly J, Wang C, Felzer B,
  Melillo JM, Jacoby HD (2009).
\newblock \enquote{Probabilistic Forecast for Twenty-First-Century Climate
  Based on Uncertainties in Emissions (Without Policy) and Climate Parameters.}
\newblock \emph{Journal of Climate}, \textbf{22}(19), 5175--5204.
\newblock \doi{10.1175/2009JCLI2863.1}.

\bibitem[{Vihola(2012)}]{vihola2012}
Vihola M (2012).
\newblock \enquote{Robust Adaptive {{Metropolis}} Algorithm with Coerced
  Acceptance Rate.}
\newblock \emph{Statistics and Computing}, \textbf{22}(5), 997--1008.
\newblock \doi{10.1007/s11222-011-9269-5}.

\bibitem[{Waggoner and Zha(1999)}]{waggoner_conditional_1999}
Waggoner DF, Zha T (1999).
\newblock \enquote{{Conditional Forecasts in Dynamic Multivariate Models}.}
\newblock \emph{The Review of Economics and Statistics}, \textbf{81}(4),
  639--651.
\newblock \doi{10.1162/003465399558508}.

\bibitem[{Wang and Woźniak(2025{\natexlab{a}})}]{WangWozniak2025}
Wang X, Woźniak T (2025{\natexlab{a}}).
\newblock \enquote{Bayesian Analyses of Structural Vector Autoregressions with
  Sign, Zero, and Narrative Restrictions Using the R Package bsvarSIGNs.}
\newblock \emph{University of Melbourne Working Paper}, pp. 1--21.
\newblock \doi{10.48550/arXiv.2501.16711}.

\bibitem[{Wang and Woźniak(2025{\natexlab{b}})}]{bsvarSIGNs}
Wang X, Woźniak T (2025{\natexlab{b}}).
\newblock \emph{bsvarSIGNs: Bayesian SVARs with Sign, Zero, and Narrative
  Restrictions}.
\newblock \doi{10.32614/CRAN.package.bsvarSIGNs}.
\newblock R package version 2.0,
  \urlprefix\url{https://CRAN.R-project.org/package=bsvarSIGNs}.

\bibitem[{Wo{\'z}niak(2016)}]{wozniak2016}
Wo{\'z}niak T (2016).
\newblock \enquote{Bayesian {{Vector Autoregressions}}.}
\newblock \emph{Australian Economic Review}, \textbf{49}(3), 365--380.
\newblock \doi{10.1111/1467-8462.12179}.

\bibitem[{Woźniak(2024{\natexlab{a}})}]{bsvars}
Woźniak T (2024{\natexlab{a}}).
\newblock \emph{bsvars: Bayesian Estimation of Structural Vector Autoregressive
  Models}.
\newblock \doi{10.32614/CRAN.package.bsvars}.
\newblock R package version 3.2,
  \urlprefix\url{https://CRAN.R-project.org/package=bsvars}.

\bibitem[{Woźniak(2024{\natexlab{b}})}]{wozniak2024}
Woźniak T (2024{\natexlab{b}}).
\newblock \enquote{Fast and Efficient Bayesian Analysis of Structural Vector
  Autoregressions Using the R Package bsvars.}
\newblock \emph{University of Melbourne Working Paper}, pp. 1--25.
\newblock \doi{10.48550/arXiv.2410.15090}.

\bibitem[{Woźniak(2025)}]{bpvars}
Woźniak T (2025).
\newblock \emph{bpvars: Forecasting with Bayesian Panel Vector
  Autoregressions}.
\newblock R package version 1.0, \urlprefix\url{http://bsvars.org/bpvars/}.

\bibitem[{Yu \emph{et~al.}(2023)Yu, Ševčíková, Raftery, and
  Curran}]{yu_probabilistic_2023}
Yu CC, Ševčíková H, Raftery AE, Curran SR (2023).
\newblock \enquote{Probabilistic County-Level Population Projections.}
\newblock \emph{Demography}, \textbf{60}(3), 915--937.
\newblock \doi{10.1215/00703370-10772782}.

\bibitem[{Zellner and Hong(1989)}]{zellner_hong}
Zellner A, Hong C (1989).
\newblock \enquote{Forecasting international growth rates using Bayesian
  shrinkage and other procedures.}
\newblock \emph{Journal of Econometrics}, \textbf{40}(1).
\newblock \doi{10.1016/0304-4076(89)90036-5}.

\end{thebibliography}

\end{document}